\documentclass[a4paper]{amsart}
\RequirePackage{amsmath}
\RequirePackage{bm}
\RequirePackage{amssymb}
\RequirePackage{upref}
\RequirePackage{amsthm}
\RequirePackage{enumerate}
\RequirePackage{pb-diagram}
\RequirePackage{amsfonts}
\RequirePackage[mathscr]{eucal}
\RequirePackage{verbatim}
\RequirePackage{xr}
\RequirePackage{graphicx}
\usepackage{calc}
\usepackage{xspace}
\RequirePackage{color}
\RequirePackage{ifthen}

\newcommand{\cf}{cf.\@\xspace}
\newcommand{\resp}{resp.\@\xspace}

\newcommand{\al}{\alpha}
\newcommand{\bet}{\beta}
\newcommand{\ga}{\gamma}
\newcommand{\de}{\delta }
\newcommand{\e}{\epsilon}

\newcommand{\f}{\varphi}
\newcommand{\h}{\eta}

\newcommand{\ka}{\kappa}
\newcommand{\lam}{\lambda}
\newcommand{\m}{\mu}

\newcommand{\om}{\omega}

\newcommand{\vt}{\vartheta}

\newcommand{\s}{\sigma}

\newcommand{\z}{\zeta}

\newcommand{\C}{\varGamma}
\newcommand{\D}{\varDelta}
\newcommand{\F}{\varPhi}
\newcommand{\Lam}{\varLambda}
\newcommand{\Om}{\varOmega}

\newcommand{\fv}[2]{#1\hspace{0pt}_{|_{#2}}}

\newcommand{\so}{{\mc S_0}}

\newcommand{\const}{\tup{const}}

\newcommand{\msp[1]}[1]{\mspace{#1mu}}

\newcommand{\R}[1][n+1]{{\protect\mathbb R}^{#1}}
\newcommand{\Ccc}[1][n]{{\protect\mathbb C}^{#1}}

\newcommand{\Cc}{{\protect\mathbb C}}

\newcommand{\N}{{\protect\mathbb N}}

\newcommand{\Z}{{\protect\mathbb Z}}
\newcommand{\eR}{\stackrel{\lower1ex \hbox{\rule{6.5pt}{0.5pt}}}{\msp[3]\R[]}}
\newcommand{\eN}{\stackrel{\lower1ex \hbox{\rule{6.5pt}{0.5pt}}}{\msp[1]\N}}
\newcommand{\eO}{\stackrel{\lower1ex \hbox{\rule{6pt}{0.5pt}}}{\msc O}}
\newcommand{\mf}[1]{\mathfrak {#1}}

\DeclareMathOperator{\supp}{supp}

\DeclareMathOperator{\Ad}{Ad}

\DeclareMathOperator{\I}{I}

\DeclareMathOperator{\tr}{tr}
\DeclareMathOperator{\SO}{SO}
\DeclareMathOperator{\SU}{SU}
\DeclareMathOperator{\U}{U}
\DeclareMathOperator{\soc}{\mf{so}}
\DeclareMathOperator{\suc}{\mf{su}}
\DeclareMathOperator{\uc}{\mf{u}}
\DeclareMathOperator{\diag}{diag}

\newcommand\ra{\rightarrow}

\newcommand\hra{\hookrightarrow}

\newcommand{\rha}{\rightharpoondown}

\newcommand\pa{\partial}
\newcommand\pde[2]{\frac {\partial#1}{\partial#2}}
\newcommand{\un}{\infty}
\newcommand{\A}{\forall}

\newcommand{\set}[2]{\{\,#1\colon #2\,\}}
\newcommand{\uu}{\cup}
\newcommand{\ii}{\cap}
\newcommand{\uuu}{\bigcup}

\newcommand{\uud}{ \stackrel{\lower 1ex \hbox {.}}{\uu}}
\newcommand{\uuud}[1]{ \stackrel{\lower 1ex \hbox {.}}{\uuu_{#1}}}
\newcommand\su{\subset}

\newcommand\eS{\emptyset}
\newcommand{\sminus}[1][28]{\raise 0.#1ex\hbox{$\scriptstyle\setminus$}}

\newcommand{\wed}{\wedge}

\newcommand\ti{\times }

\newcommand{\abs}[1]{\lvert#1\rvert}

\newcommand{\norm}[1]{\lVert#1\rVert}

\newcommand{\spd}[2]{\protect\langle #1,#2\protect\rangle}

\newcommand\cha[3]{{\bar\varGamma}_{#1#2}^#3}

\newcommand{\tit}{\textit}

\newcommand{\tup}{\textup}% text upright

\newcommand{\mc}{\protect\mathcal}
\newcommand{\msc}{\protect\mathscr}

\providecommand{\bysame}{\makebox[3em]{\hrulefill}\thinspace}

\newcommand{\cq}[1]{\glqq{#1}\grqq\,}

\newcommand{\bt}{\begin{thm}}
\newcommand{\bl}{\begin{lem}}
\newcommand{\bc}{\begin{cor}}
\newcommand{\bd}{\begin{definition}}
\newcommand{\bpp}{\begin{prop}}
\newcommand{\br}{\begin{rem}}
\newcommand{\bn}{\begin{note}}
\newcommand{\be}{\begin{ex}}
\newcommand{\bes}{\begin{exs}}
\newcommand{\bb}{\begin{example}}
\newcommand{\bbs}{\begin{examples}}
\newcommand{\ba}{\begin{axiom}}
\newcommand{\bas}{\begin{assumption}}

\newcommand{\et}{\end{thm}}
\newcommand{\el}{\end{lem}}
\newcommand{\ec}{\end{cor}}
\newcommand{\ed}{\end{definition}}
\newcommand{\epp}{\end{prop}}
\newcommand{\er}{\end{rem}}
\newcommand{\en}{\end{note}}
\newcommand{\ee}{\end{ex}}
\newcommand{\ees}{\end{exs}}
\newcommand{\eb}{\end{example}}
\newcommand{\ebs}{\end{examples}}
\newcommand{\ea}{\end{axiom}}
\newcommand{\eas}{\end{assumption}}

\newcommand{\bp}{\begin{proof}}
\newcommand{\ep}{\end{proof}}
\newcommand{\eps}{\renewcommand{\qed}{}\end{proof}}

\newcommand{\bal}{\begin{align}}

\newcommand{\bi}[1][1.]{\begin{enumerate}[\upshape #1]}
\newcommand{\bia}[1][(1)]{\begin{enumerate}[\upshape #1]}
\newcommand{\bin}[1][1]{\begin{enumerate}[\upshape\bfseries #1]}
\newcommand{\bir}[1][(i)]{\begin{enumerate}[\upshape #1]}
\newcommand{\bic}[1][(i)]{\begin{enumerate}[\upshape\hspace{2\cma}#1]}
\newcommand{\bis}[2][1.]{\begin{enumerate}[\upshape\hspace{#2\parindent}#1]}
\newcommand{\ei}{\end{enumerate}}

\newcommand\ndots{\raise 0.47ex \hbox {,}\hskip0.06em\cdots %
     \raise 0.47ex \hbox {,}\hskip0.06em} 

\newcommand{\q}{\quad}
\newcommand{\qq}{\qquad}

\newcommand{\hp}{\hphantom}

\newcommand\nd{\noindent}

\newskip\Csmallskipamount                                                
\Csmallskipamount=\smallskipamount
\newskip\Cmedskipamount
\Cmedskipamount=\medskipamount
\newskip\Cbigskipamount
\Cbigskipamount=\bigskipamount

\newcommand\cvs{\vspace\Csmallskipamount}   
\newcommand\cvm{\vspace\Cmedskipamount}

\newskip\csa
\csa=\smallskipamount

\newskip\cma
\cma=\medskipamount

\newskip\cba
\cba=\bigskipamount

\newdimen\spt
\newcommand\citem{\cvs\advance\itemno by
1{(\romannumeral\the\itemno})\hskip3pt}
\newcommand{\bitem}{\cvm\nd\advance\itemno by
1{\bf\the\itemno}\hspace{\cma}}

\newcount\itemno
\newcommand{\las}[1]{\label{S:#1}}

\newcommand{\lae}[1]{\label{E:#1}}
\newcommand{\lat}[1]{\label{T:#1}}
\newcommand{\lal}[1]{\label{L:#1}}

\newcommand{\lar}[1]{\label{R:#1}}

\newcommand{\rs}[1]{Section~\ref{S:#1}}

\newcommand{\rt}[1]{Theorem~\ref{T:#1}}
\newcommand{\rl}[1]{Lemma~\ref{L:#1}}

\newcommand{\re}[1]{\eqref{E:#1}}
\newcommand{\frt}[1]{Theorem~\ref{T:#1} on page~\tup{\pageref{T:#1}}}
\newcommand{\frl}[1]{Lemma~\ref{L:#1} on page~\tup{\pageref{L:#1}}}
\newcommand{\frr}[1]{Remark~\ref{R:#1} on page~\tup{\pageref{R:#1}}}

\newcommand{\fre}[1]{\eqref{E:#1} on page~\tup{\pageref{E:#1}}}

\newskip\thmskip
\thmskip=\parindent

\newskip\hsk
\newenvironment{hinw}{\labelsep=0pt\begin{list}{}{\labelsep=0pt\itemindent=0pt\labelwidth=0pt\leftmargin=\parindent\rightmargin=0pt\partopsep=\cba}%
\item\it\nopagebreak\nopagebreak}%
{\end{list}}

\newcommand\bh{\begin{hinw}}
\newcommand{\eh}{\end{hinw}}

\newtheoremstyle{normal}% name
  {\cba}%      Space above, empty = `usual value'
  {\cba}%      Space below
  {}% Body font
  {\thmskip}%Indent amount (empty = no indent, \parindent = para indent)
  {\bfseries}% Thm head font
  {.}%        Punctuation after thm head
  {\hsk}%     Space after thm head: " " = normal interword space;
  {}% Thm head spec

\newtheoremstyle{abschnitt}% name
  {\cba}%      Space above, empty = `usual value'
  {\cba}%      Space below
  {}% Body font
  {\thmskip}% Indent amount (empty = no indent, \parindent = para indent)
  {\bfseries}% Thm head font
  {.}%        Punctuation after thm head
  {\hsk}%     Space after thm head: " " = normal interword space;
  {}% Thm head spec

\newtheoremstyle{italic}% name
  {\cba}%      Space above, empty = `usual value'
  {\cba}%      Space below
  {\itshape}% Body font
  {\thmskip}%  Indent amount (empty = no indent, \parindent = para indent)
  {\bfseries}% Thm head font
  {.}%        Punctuation after thm head
  {\hsk}%     Space after thm head: " " = normal interword space;
  {}% Thm head spec

\newtheoremstyle{aufgaben}% name
  {\cba}%      Space above, empty = `usual value'
  {\cba}%      Space below
  {}% Body font
  {}%         Indent amount (empty = no indent, \parindent = para indent)
  {\normalsize\bfseries}% Thm head font
  {.}%        Punctuation after thm head
  {\hsk}%     Space after thm head: " " = normal interword space;
  {}% Thm head spec

\newtheoremstyle{break}% name
  {\cba}%      Space above, empty = `usual value'
  {\cba}%      Space below
  {\itshape}% Body font
  {}%         Indent amount (empty = no indent, \parindent = para indent)
  {\bfseries}% Thm head font
  {.}%        Punctuation after thm head
  {\newline}% Space after thm head: \newline = linebreak
  {}%         Thm head spec

\swapnumbers
\theoremstyle{italic}
\newtheorem{thm}[subsection]{Theorem}
\newtheorem{lem}[subsection]{Lemma}
\newtheorem{prop}[subsection]{Proposition}
\newtheorem{cor}[subsection]{Corollary}

\theoremstyle{normal}
\newtheorem{rem}[subsection]{Remark}
\newtheorem{definition}[subsection]{Definition}
\newtheorem{example}[subsection]{Example}
\newtheorem{examples}[subsection]{Examples}
\newtheorem{ex}[subsection]{Exercise}
\newtheorem{note}[subsection]{}
\newtheorem{axiom}[subsection]{Axiom}
\newtheorem{assumption}[subsection]{Assumption}

\theoremstyle{aufgaben}
\newtheorem{exs}[subsection]{Exercises}

\swapnumbers

\numberwithin{equation}{section}
\numberwithin{figure}{section}

\newenvironment{textequation}[1][0.8]
{\begin{equation}
\begin{aligned}
\begin{minipage}{#1\linewidth}}
{\end{minipage}
\end{aligned}
\end{equation}
\ignorespacesafterend}

\newcommand{\btext}{\begin{textequation}}
\newcommand{\etext}{\end{textequation}}

\def\hinweis{\@startsection{subsection}{2}%
 \z@{0.7\linespacing\@plus 0.5\linespacing}{0.7\linespacing}%
{\normalfont\itshape\indent}}

\newcounter{hours}\newcounter{minutes}
\newcommand{\printtime}{%
\setcounter{hours}{\time/60}%
\setcounter{minutes}{\time-\value{hours}*60}%
\ifthenelse{\value{minutes}<9}{\thehours :0\theminutes}{\thehours:\theminutes}}

\usepackage[german,english]{babel}
\usepackage{graphicx}
\RequirePackage{amsmath}
\RequirePackage{bm}
\RequirePackage{amssymb}
\RequirePackage{upref}
\RequirePackage{amsthm}
\RequirePackage{enumerate}
\RequirePackage{pb-diagram}
\RequirePackage{amsfonts}
\RequirePackage[mathscr]{eucal}
\RequirePackage{verbatim}
\RequirePackage{xr}
\RequirePackage{graphicx}
\usepackage{calc}
\usepackage{xspace}
\makeatletter
\RequirePackage{color}
\newcommand{\ann}[1]{\renewcommand{\@makefnmark}{\mbox{$^{\color{red}{\@thefnmark}}$}}%
\footnote {#1}}
\makeatother

\RequirePackage{upref}
\RequirePackage{amsthm}
\RequirePackage{enumerate}%\begin{enumerate}[(i)]
\usepackage[mathscr]{eucal}
\usepackage{xr-hyper}

\usepackage{calc}

\newlength{\oddsidemarginlength}
\newlength{\topmarginlength}

\newcounter{numberoflines}
\newcounter{tempcc}
\oddsidemargin=\oddsidemarginlength
\evensidemargin=\oddsidemargin
\topmargin=\topmarginlength
\usepackage[colorlinks=true,linkcolor=blue,citecolor=blue,urlcolor=blue]{hyperref}  

\begin{document}

\flushbottom

%\larger[1]
%\frontmatter

\title[Combining gravity with the standard model]{Combining gravity with the forces of the standard model on a cosmological scale}

% author one information
\author{Claus Gerhardt}
\address{Ruprecht-Karls-Universit\"at, Institut f\"ur Angewandte Mathematik,
Im Neuenheimer Feld 294, 69120 Heidelberg, Germany}
%\curraddr{}
\email{gerhardt@math.uni-heidelberg.de}
\urladdr{http://web.me.com/gerhardt/}
%\thanks{This work has been supported by the DFG}

% author two information
%\author{}
%\address{}
%\curraddr{}
%\email{}
%\thanks{}
%
%\PACS{98.80.Qc}
\date{\today}
%
% at present the "communicated by" line appears only in ERA and PROC
%\commby{}

%\dedicatory{}

\begin{abstract}
We prove the existence of a spectral resolution of the Wheeler-DeWitt equation when the underlying spacetime is a Friedman universe with flat spatial slices and where the matter fields are comprised of the strong interaction, with $\SU(3)$  replaced by a general $\SU(n)$, $n\ge 2$, and the electro-weak interaction.

The wave functions are maps from $\R[4n+10]$ to a subspace of the antisymmetric Fock space, and one noteworthy result is that, whenever the electro-weak interaction is involved, the image of an eigenfunction is in general not one dimensional, i.e.,  in general it makes no sense  specifying a fermion and looking for an eigenfunction the range of which is contained in the one dimensional vector space spanned by the fermion.
\end{abstract}

\maketitle

\tableofcontents

\setcounter{section}{0}
\section{Introduction}
In three former papers \cite{cg:qfriedman,cg:qfriedman-ym,cg:qfriedman-ym2} we proved a spectral resolution of the Wheeler-DeWitt equation in the cosmological case---at least in principle. When the spatial slices of the underlying Friedman-Robertson-Walker universe are flat we have developed a model in \cite{cg:qfriedman-ym2} with strictly positive energy levels---albeit for a single $\SO(3)$ gauge field. For a definition of positive energy levels in this situation see \cite[introduction]{cg:qfriedman-ym2}.

In Friedman-Robertson-Walker models the matter Lagrangians must reflect the spacetime symmetries up to  gauge transformations, and hence very special ans\"atze for the gauge fields have to be considered. For $\SO(n)$ \resp $\SU(n)$ gauge fields such ans\"atze are known for some time, \cf  \cite{bertolami:gauge} and \cite{moniz:gauge}, but due to their special nature these ans\"atze introduce a number of non-dynamical variables into the Lagrangian resulting in additional first-class constraints. Hence, any attempt to generalize our previous results to higher dimensional gauge groups faced two major challenges, first, to handle these additional constraints and second, to handle a large number of dynamical bosonic variables---in fact any number larger than $1$ posed a problem for the actual spectral resolution when an implicit eigenvalue problem for the gravitational Hamiltonian has to be solved and one has to prove that a (weighted) $L^2$-norm is compact compared with the gravitational energy norm. The former proof only worked in case of  a single bosonic matter variable.

These difficulties could be solved: the additional constraint equations are taken care of by considering a special infinite dimensional subspace 
\begin{equation}\lae{1.1}
E\su C^\un_c(\R[4n+10],\mc F),
\end{equation}
where $\mc F$ is a finite dimensional subspace of the antisymmetric Fock space, as the core domain, while in case of the implicit eigenvalue problem the compactness property could be proved.

We consider as underlying spacetime a Friedman-Robertson-Walker space $N=N^4$ with flat spatial sections and the Lagrangian functional has the form
\begin{equation}
J=\al_M^{-1}\int_\Om(\bar R-2\Lam)+\int_\Om L_{M_1}+\int_\Om L_{M_2},
\end{equation}
where $L_{M_1}$ is the Lagrangian of the strong interaction, though we have replaced the $SU(3)$ connection by a general $\SU(n)$, $n\ge 2$, connection, and $L_{M_2}$ is the Lagrangian for the electro-weak interaction.

The cosmological constant $\Lam$ is very important, since it will play the role of an eigenvalue when we solve the implicit eigenvalue problem. It will turn out that $\Lam$ has to be negative.

The core domain $E$ in \re{1.1} can be written as an orthogonal sum
\begin{equation}
E=\bigoplus_{1\le k,l\le 9}E_{kl},
\end{equation}
where 
\begin{equation}
E_{kl}\su C^\un_c(\R[4n+10], F_{\s_k}\otimes F_{\rho_l}) 
\end{equation}
and $F_{\s_k}$ \resp $F_{\rho_l}$ are orthogonal subspaces in the fermion spaces $\mc F_1$ \resp $\mc F_2$ spanned by the fermions of the strong \resp electro-weak interaction. For the electro-weak interaction we have
\begin{equation}
\mc F_2=\bigoplus_{1\le l\le 9}F_{\rho_l},
\end{equation}
but the $F_{\s_k}$ fail to generate $\mc F_1$. Each of the $E_{kl}$ generates an infinite dimensional Hilbert space $\mc H_{kl}$ in which we solve a spectral resolution for the Wheeler-deWitt equation. Since the $\mc H_{kl}$ are mutually orthogonal we can then define a spectral resolution in the orthogonal sum.

The main results can be summarized in:
\bt
There exist $81$ Hilbert spaces $\mc H_{kl}$ as described above, a detailed description will be given in the last three sections, and a self-adjoint operator $H$ in
\begin{equation}
\mc H=\bigoplus_{1\le k,l\le 9}\mc H_{kl},
\end{equation}
such that, for fixed $(k,l)$, there exists a complete sequence of eigenfunctions $\tilde\Psi_{ij}\in \mc H_{kl}$, $(i,j)\in\N\times\N$, with eigenvalues $\lam_{ij}$ of finite multiplicities satisfying
\begin{equation}
H\tilde\Psi_{ij}=\lam_{ij}\tilde \Psi_{ij},
\end{equation}
\begin{equation}
0<\lam_{ij}\q\wed\q \lim_{i\ra\un}\lam_{ij}=\un\q\wed\q\lim_{j\ra\un}\lam_{ij}=0.
\end{equation}
The eigenfunctions are maps from
\begin{equation}
\tilde\Psi_{ij}:\R[4n+10]\ra F_{\s_k}\otimes F_{\rho_l}.
\end{equation}
Let $t$ be the variable which corresponds to the logarithm of the scale factor, then the rescaled eigenfunctions
\begin{equation}
\Psi_{ij}(t,\cdot)=\tilde\Psi_{ij} (t-\tfrac12\log \lam_{ij},\cdot)
\end{equation}
are solutions of the Wheeler-DeWitt equation with cosmological constant
\begin{equation}
\Lam_{ij}=-\lam_{ij}^{-3}.
\end{equation}
\et
\br
(i) Instead  of considering both the strong and the electro-weak interactions each can be treated separately leading to similar results.

\cvm
(ii) The method of proof can be applied to finitely many matter fields.

\cvm
(iii) Whenever the electro-weak interaction is involved the eigenfunctions $\Psi$ in general  cannot be written as simple products
\begin{equation}
\Psi=u\h,
\end{equation}
such that
\begin{equation}
\h\in \mc F_1\otimes\mc F_2\q\wed\q u(x)\in \Cc\qq\A\, x\in \R[4n+10].
\end{equation}
Thus, in general it makes no sense specifying a fermion $\h$ and looking for an eigenfunction $\Psi$ satisfying
\begin{equation}
R(\Psi)\su \langle\h\rangle.
\end{equation}

\cvm
(iv) The number $81$ of mutually orthogonal Hilbert spaces is due to the fact that the fermionic constraint operators $\hat l_k$ \resp $\hat\lam_0$ of the strong ($\SU(n)$) \resp electro-weak interaction each have exactly $9$ eigenspaces due to their definitions as the sum of number operators.
\er
\section{Conventions and definitions}\las{2}
In this section we give a brief overview of our conventions and definitions. 

We denote the Minkowski metric by $\h_{ab}$, $0\le a,b\le 3$,
\begin{equation}
(\h_{ab})=\diag(-1,1,1,1)
\end{equation}
and define the Dirac matrices accordingly
\begin{equation}\lae{2.2}
\{\ga^a,\ga^b\}_+=2\h^{ab}.
\end{equation}
$\ga^0$ is antihermitean and $\ga^k$ hermitean. When we are dealing with normal spinors, e.g., in case of the strong interaction,  we choose a basis such that
\begin{equation}\lae{2.3}
\ga^0=i\begin{pmatrix}
\I&0\\[\cma]
0&-\I
\end{pmatrix}.
\end{equation}

However, when Weyl spinors are considered, e.g., in case of the electro-weak interaction,  we choose a basis such that the helicity operator $\ga^5$ is represented as 
\begin{equation}
\ga^5=-\ga^0\ga^1\ga^2\ga^3=i\begin{pmatrix}
\I&0\\[\cma]
0&-\I
\end{pmatrix},
\end{equation}
then $\ga^0$ has the form
\begin{equation}
\ga^0=i\begin{pmatrix}
0&\I\\[\cma]
\I&0
\end{pmatrix}.
\end{equation}
The $\ga^k$, $1\le k\le 3$, are defined by 
\begin{equation}\lae{2.6}
\ga^k=i\begin{pmatrix}
0&\s_k\\[\cma]
-\s_k&0
\end{pmatrix}
\end{equation}
in both cases, where $\s_k$ are the Pauli matrices.

Let $\psi=(\psi_a)$ be a spinor, then a bar simply denotes complex conjugation
\begin{equation}
\bar\psi=(\bar\psi_a);
\end{equation}
the symbol $\tilde\psi$ is defined by
\begin{equation}
\tilde\psi= i\bar\psi\ga^0,
\end{equation}
where the notation on the right-hand side automatically implies that now $\bar\psi$ has to be understood as a row, since $\ga^0$ acts from the right. 

The meaning of symbols  may depend on the section where they are used, e.g., the symbols $\norm\cdot$  \resp $\norm\cdot_1$ denote different norms, though their specific definitions will depend on the contexts in which they are used, though $\norm\cdot$ always denotes a (weighted) $L^2$-norm and $\norm\cdot_1$ a stronger energy norm.

Let $\Om\su \R[n]$, $1\le n$, be an open set, then we denote by
\begin{equation}
H^{1,2}(\Om)
\end{equation}
the usual Sobolev space with norm
\begin{equation}
\int_\Om\{\abs{D u}^2+\abs u^2\}.
\end{equation}

When $E$ is Banach space and $\Om\su\R[n]$ as before we denote the space of test functions defined in $\Om$ with values in $E$ by
\begin{equation}
C^\un_c(\Om,E).
\end{equation}

We also use a correction term $\chi_0$ occasionally  when defining the Lagrangian, which is a \tit{function} defined in the space of Lorentz metrics on $N$ such that, when $\chi_0$ is evaluated at a metric of the form
\begin{equation}
d\bar s^2=-w^{-2}(dx^0)^2+e^{2f}\s_{ij}dx^idx^j,
\end{equation}
then
\begin{equation}
\chi_0=e^{6f},
\end{equation}
\cf \cite[Lemma 3.1]{cg:qfriedman-ym}.
\section{The strong interaction}\las{3}
The underlying gauge group for the strong interaction is $\SU(3)$. We shall consider a general $\SU(n)$, $n\ge 2$, instead, since an arbitrary $n$ poses no greater challenges.

As already mentioned in the introduction we have to look at very special gauge fields that reflect the symmetries of the underlying spacetime up to a gauge transformation. When the spacetime is a Friedman-Robertson-Walker space  which is topologically either
\begin{equation}
N=\R[]\ti S^3
\end{equation}
or
\begin{equation}
N=\R[]\ti \R[3]
\end{equation}
the gauge fields have to be either $\SO(4)$ symmetric, i.e., symmetric with respect to both left and right actions of $\SU(2)\cong SO(3)$ on the spacelike sections of $N$, or symmetric with respect to rigid motions in $\R[3]$ after an appropriate gauge transformation.

Let the spacetime metric satisfy
\begin{equation}\lae{3.3}
d\bar s^2=-w^2d{x^0}^2+e^{2f}\s_{ij}dx^idx^j,
\end{equation}
where $(\s_{ij})$ is the standard metric of a space of constant curvature $\so$, at the moment we allow the possibilities $\so=S^3$ or $\so=\R[3]$, but later we shall stipulate $\so=\R[3]$, and let the left-invariant $1$-forms $\om^a$, $1\le a\le 3$ satisfy
\begin{equation}\lae{3.4}
\s_{ij}=\de_{ab}\om^a_i\om^b_j\q\wed\q \s^{ij}\om^a_i\om^b_j=\de^{ab}
\end{equation}
and
\begin{equation}\lae{3.5}
d\om^a=
\begin{cases}
0,&\so=\R[3],\\
\tfrac12\e^a_{\hp{a}bc}\om^b\om^c,&\so=S^3.
\end{cases}
\end{equation}
Let $E_{km}$ be the matrices
\begin{equation}\lae{3.6}
E_{km}=(\de^i_k\de_{mj})
\end{equation}
for $1\le k,m\le n+3$ and set
\begin{equation}\lae{3.7}
T_{km}=E_{km}-E_{mk}
\end{equation}
for $1\le k\ne m\le n+3$.

The $T_{km}$ with $1\le k<m\le 3$ are generators of $\soc(3)$ or equivalently of the Lie algebra of the adjoint representation of $\SU(2)$ which is isomorphic to $\suc(2)$.  The precise correspondence with the Pauli matrices will be given later in \rs{6}. 

We stipulate that the indices $a,b,c$, when used in connection with these generators or with the matrices in \re{3.6} or \re{3.7}, will always run from $1$ to $3$.

Following \cite{bertolami:gauge}\footnote{In the appendix of this paper the necessary procedures for a spacetime $N=\R[]\ti \so$ with a general homogeneous space $\so$ is described.} and \cite{moniz:gauge} we define the connection $A=A(t)$ by
\begin{equation}\lae{3.8}
A(t)=\hat A(t)+B(t),
\end{equation}
where
\begin{equation}
\hat A(t)=(\Lam^{km}(t)E_{k+3,m+3}-\tfrac13\Lam^k_k(t)E^a_a)dt,
\end{equation}
\begin{equation}
\begin{aligned}
B(t)=(-\f_0T_{bc}\e_a^{\hp{a}bc} +\bar z^k(t)E_{a,k+3}-z^kE_{k+3,a})\om^a_idx^i,
\end{aligned}
\end{equation}
$(\Lam^{km}(t))$, $1\le k,m\le n$, is an arbitrary antihermitian matrix, $\f_0=\f_0(t)$ a real function and $z^k=z^k(t)$, $1\le k\le n$, arbitrary complex valued functions. The bar indicates complex conjugation.

Writing
\begin{equation}
A=A_\mu dx^\mu
\end{equation}
the connection $(A_\mu)$ then has values in $\suc(n+3)$. The connection
\begin{equation}
\hat A=\hat A_\mu dx^\mu=\hat A_0dx^0
\end{equation}
can be viewed as being a general element of $\uc(n)$, when $\hat A_0$ is considered to be a homomorphism in the $n$-dimensional subspace of $\Ccc[n+3]$ defined by
\begin{equation}\lae{3.13}
\set{\zeta=(0,0,0,\zeta^{1+3},\ldots,\zeta^{n+3})}{\zeta^{k+3}\in\Cc,\; 1\le k\le n}\cong \Ccc.
\end{equation}
For convenience we shall label the components of $\zeta$ in the form
\begin{equation}\lae{3.14}
\zeta=(0,0,0,\zeta^k)\equiv (\zeta^k)
\end{equation}
in this case.

However, we shall consider $\hat A_0$ as a general $\U(n)$ connection only for $n=1$. In case $n\ge 2$ we shall in addition require
\begin{equation}
\Lam^k_k=0
\end{equation}
such that $A_0$ has values in $\suc(n)$. $\hat A=\hat A(t)$ will then be the actual $\SU(n)$ connection.

The corresponding matter Lagrangian comprises three terms: the energy of the gauge field
\begin{equation}
L_{{Y\mspace{-3mu}M}_1} =\tfrac14\tr(F_{\mu\lam} F^{\mu\lam}),
\end{equation}
a Higgs term
\begin{equation}\lae{3.17}
L_{H_1}=-\big(\tfrac12 \bar g^{\mu\lam}\F_\mu\bar\F_\lam\chi_0^{-\frac13}+U(\F)\chi_0^{-\frac23}\big),
\end{equation}
and a massive Dirac Lagrangian describing the fermionic sector 
\begin{equation}\lae{3.18}
L_{F_1}=-\tfrac12\{\tilde\psi_iE^\mu_a\ga^a(D_\mu \psi)^i+\overline{\tilde\psi_iE^\mu_a\ga^a(D_\mu \psi)^i}\} -m\tilde\psi_i\psi^i\chi_0^{-\frac16}. 
\end{equation}

\bl\lal{3.1}
Let $\so=\R[3]$ and $A$ be the connection in \re{3.8}, then its energy 
\begin{equation}
F^2=-\tr(F_{\mu\lam}F^{\mu\lam})
\end{equation}
can be expressed as
\begin{equation}
\begin{aligned}\lae{3.20}
F^2=-12\{2\abs{\dot\f_0}^2+\abs{\tfrac D{dt}z}^2\}w^{-2}e^{-2f}+12\{\f_0^4+8\f_0^2\abs z^2+\abs z^4\}e^{-4f},
\end{aligned}
\end{equation}
where, in case $n\ge 2$,
\begin{equation}\lae{3.21}
\tfrac D{dt}z^k=\dot z^k+\Lam^k_mz^m,
\end{equation}
and $\Lam\in \suc(n)$, while for $n=1$, $\Lam\in \uc(1)$,
\begin{equation}
\Lam=\Lam^{11}=i\vt(t),\qq \vt(t)\in\R[],
\end{equation}
and
\begin{equation}\lae{3.23}
\tfrac D{dt}z=\dot z+\tfrac43 i \vt z.
\end{equation}
\el
\bp
The proof is  straight-forward by observing that, when choosing local coordinates such that $\om^a_j=\de^a_j$,
\begin{equation}
\begin{aligned}
F_{0j}=-\dot\f_0\e_a^{\hp a bc}T_{bc}\om^a_j+\{-\tfrac D{dt}z^kE_{k+3,j}+\overline{\tfrac D{dt} z^m} E_{j,m+3}\},
\end{aligned}
\end{equation}
where the different definitions of the covariant derivative of $z$ is due to the fact that, in case $n\ge 2$, $\Lam$ has the trace zero.

The other non-vanishing components $F_{ij}$, $i\ne j$, are
\begin{equation}
\begin{aligned}
F_{ij}=& -4\f_0^2\e_i^{\hp i bc}\e_{jb}^{\hp{jb}c'}T_{cc'}-4\f_0\bar z^k\e_i^{\hp ibj}E_{b,k+3}\\
&\, +4\f_0z^k\e_i^{\hp icj}E_{k+3,c}-\abs z^2T_{ij}.
\end{aligned}
\end{equation}
The final result is then a simple computation.
\ep

Let us now look at the Higgs term. The scalar field $\F=(\F^k)$ has values in $\Ccc[n+3]$, or effectively in $\Ccc$, according to the conventions in \re{3.13} and \re{3.14}.

The covariant derivative $D_\mu\F=\F_\mu$ can be defined either by
\begin{equation}\lae{3.26}
\F_\mu=\F_{,\mu}+g_1A_\mu\F
\end{equation}
or by
\begin{equation}\lae{3.27}
\F_\mu=\F_{,\mu}+g_1\hat A_\mu\F,
\end{equation}
where $g_1$ is a positive coupling constant. Both definitions make sense. In \re{3.26} we consider the full connection $A$, while in \re{3.27} only the effective connection $\hat A\in \suc(n)$ \resp $\hat A\in \uc(1)$, when $n=1$, is taken into account.

Evaluating
\begin{equation}
\abs{D\F}^2=\bar g^{\mu\lam}\F_\mu\bar\F_\lam
\end{equation}
in case of \re{3.26} we obtain
\begin{equation}\lae{3.29}
\abs{D\F}^2=-w^{-2}\abs{\tfrac D{dt}\F}^2+3g_1^2e^{-2f}\abs{\spd{\F}z}^2,
\end{equation}
where
\begin{equation}\lae{3.30}
\tfrac D{dt}\F^k=\dot\F^k+g_1\Lam^k_m\F^m
\end{equation}
and
\begin{equation}
\spd{\F}z=\F_k\bar z^k.
\end{equation}
In case of \re{3.27} we have
\begin{equation}
\abs{D\F}^2=-w^{-2}\abs{\tfrac D{dt}\F}^2.
\end{equation}

The additional lower order term in \re{3.29} would have the effect that the bosonic Hilbert space, we will be working in after quantization, would no longer be invariant with respect to the corresponding Hamiltonian. Though the overall solvability wouldn't be endangered the lacking invariance suggests that the effective connection will also be the more natural one and we shall always use the definition \re{3.27}.

The potential $U=U(\F)$ should be of the form
\begin{equation}
U=U_0(\abs{\F}^2)
\end{equation}
with a smooth $U_0$ such that after quantization the resulting Hamiltonian, combining Yang-Mills and Higgs field, is self-adjoint with a complete sequence of eigenvectors having positive eigenvalues.

Requiring the estimate
\begin{equation}\lae{3.34}
-c_2+c_1\abs{\F}^{2p}\le U(\F)\le c_1'\abs{\F}^{2p}+c_2',
\end{equation}
with $1\le p\in\N$ and  positive constants $c_1,c_1'$  and non-negative $c_2,c_2'$, will guarantee a complete set of eigenvectors. However, a finite number of eigenvalues could be negative under this very weak assumptions. A positive lower bound of the eigenvalues can be proved, if either the constant $c_2$ is small relative to $c_1$ or if $U$ satisfies the additional condition
\begin{equation}
U(\F)\ge 0.
\end{equation}
Hence, the potentials
\begin{equation}
U(\F)=\lam (\abs\F^2-\mu)^2,
\end{equation}
$\lam,\mu\in\R[]$, $\lam>0$, or
\begin{equation}
U(\F)=\lam \abs\F^4+\mu\abs\F^2
\end{equation}
with $\lam>0$ and $\mu\in\R[]$ satisfying
\begin{equation}
\abs\m<c_0(\lam),
\end{equation}
would lead to positive energy levels, see \frt{9.3}.

As we already mentioned in the \rs{2} the energy $\abs{D\F}^2$ as well as the potential $U$ should be multiplied by appropriate powers of a correction term $\chi_0$ which will ensure that these terms are equipped with the right powers of the scale factor, \cf \cite[Lemma 3.1]{cg:qfriedman-ym} for details.

It turns out that $\abs{D\F}^2$ has to be multiplied by $\chi_0^{-\frac13}$ and $U$ by $\chi_0^{-\frac23}$.

Let us summarize these results in:
\bl\lal{3.2}
Choosing a coordinate system such that the metric $(\bar g_{\mu\lam})$ is expressed as in \re{3.3}, then the Higgs term \re{3.17} has the form
\begin{equation}
L_{H_1}=\tfrac12 w^{-2}\abs{\tfrac D{dt}\F}^2e^{-2f}-U(\F)e^{-4f}.
\end{equation}
\el
The Lagrangian of the fermionic field is stated in \re{3.18}. Here, $\psi=(\psi^i_a)$ is a multiplet of spinors with spin $\tfrac12$; $a$ is the spinor index, $1\le a\le 4$, and $i$, $1\le i\le n$, the \tit{colour} index, where we use the convention expressed in \re{3.14}, namely,
\begin{equation}
\psi=(0,0,0,\psi^i_a)\equiv(\psi^i_a).
\end{equation}
We will also lower or raise the index $i$ with the help of the Euclidean metric $(\de_{ij})$.

Let $\C_\mu$ be the spinor connection
\begin{equation}
\C_\mu=\tfrac14 \om_{\mu\hp ba}^{\hp\mu b}\ga_b\ga^a,
\end{equation}
then the covariant derivative $D_\mu\psi$ is defined by
\begin{equation}
D_\mu\psi=\psi_{,\mu}+\C_\mu\psi+g_1A_\mu\psi.
\end{equation}
 In contrast to the previous consideration, when we looked at the Higgs term, we do not have to worry about which connection to take, the full connection $A_\mu$ or the effective connection $\hat A_\mu$. The Lagrangian will be the same in both cases this time.

Let $(e^b_\lam)$ be a $4$-bein such that
\begin{equation}
\bar g_{\mu\lam}=\h_{ab}e^a_\mu e^b_\lam,
\end{equation}
where $(\h_{ab})$ is the Minkowski metric, and let $(E^\mu_a)$ be its inverse such that
\begin{equation}
E^\mu_a=\h_{ab}\bar g^{\mu\lam}e^b_\lam,
\end{equation}
\cf \cite[p. 246]{eguchi:book}. 

The covariant derivative of $E^\al_a$ with respect to $(\bar g_{\al\bet})$ is then given by
\begin{equation}
E^\al_{a;\mu}=E^\al_{a,\mu}+\cha \mu\bet\al E^\bet_a
\end{equation}
and
\begin{equation}
\om_{\mu\hp ba}^{\hp\mu b}=E^\lam_{a;\mu}e^b_\lam=-E^\lam_ae^b_{\lam;\mu},
\end{equation}
hence
\begin{equation}
\C_\mu=\tfrac14 \om_{\mu\hp ba}^{\hp\mu b}\ga_b\ga^a=\tfrac14 E^\lam_{a;\mu}e^b_\lam \ga_b\ga^a=-\tfrac14 E^\lam_ae^b_{\lam;\mu}\ga_b\ga^a. 
\end{equation}
If we choose in \re{3.14} $\so=\R[3]$ and $\s_{ij}=\de_{ij}$ we deduce
\begin{equation}
\C_0=0
\end{equation}
and
\begin{equation}
\C_i=\tfrac12  w^{-1} \dot fe^f\ga_i\ga^0,\qq 1\le i\le 3.
\end{equation}

To simplify the presentation we will consider the connection $\hat A$ when calculating the covariant derivatives of $\psi$, since one can easily check that the final result will not be affected by this choice.

Thus we deduce
\begin{equation}
D_0\psi=\dot\psi + g_1\hat A_0\psi,
\end{equation}
\begin{equation}
D_k\psi=\C_k\psi=\tfrac12 w^{-1}\dot fe^f\ga_k\ga^0\psi,
\end{equation}
and
\begin{equation}\lae{3.52}
\begin{aligned}
\tilde\psi_iE^\mu_a\ga^a(D_\mu\psi)^i&=\bar\psi_ii\ga^0\{E^0_\mu\ga^aD_0\psi^i+E^k_a\ga^aD_k\psi^i\}\\
&= i\bar\psi_i\ga^0\{w^{-1}\ga^0(\dot\psi^i+\Lam^i_j\psi^j)\\
&\hp = \qq\q+e^{-f}\ga^k\tfrac12 w^{-1}\dot f e^f\ga_k\ga^0\psi^i\},
\end{aligned}
\end{equation}
where we used
\begin{equation}
E^\mu_0=w^{-1}\de^\mu_0\q\wed\q E^\mu_k=e^{-f}\de^\mu_k,
\end{equation}
when $\s_{ij}=\de_{ij}$.

In view of \fre{2.2} we have
\begin{equation}
\ga^k\ga_k=3\I\q\wed\q \ga^0\ga^0=-\I,
\end{equation}
hence the right-hand side of \re{3.52} is equal to
\begin{equation}
i\bar\psi_i\ga^0\{w^{-1}\ga^0(\dot\psi^i+\Lam^i_j\psi^j)+\tfrac32 w^{-1}\dot f \ga^0\psi^i\},
\end{equation}
and we deduce further, by setting
\begin{equation}
\chi=e^{\frac32 f}\psi,
\end{equation}
\begin{equation}
\begin{aligned}
\tilde\psi_iE^\mu_a\ga^a(D_\mu\psi)^i&= i\bar\chi_i\ga^0w^{-1}\ga^0\tfrac D{dt}\chi^ie^{-3f}\\
&=-i\bar\chi_i\tfrac D{dt}\chi^iw^{-1}e^{-3f},
\end{aligned}
\end{equation}
where
\begin{equation}\lae{3.58}
\tfrac D{dt}\chi^i=\dot\chi^i+g_1\Lam^i_j\chi^j.
\end{equation}

Summarizing the preceding results we obtain:
\bl\lal{3.3}
The Dirac Lagrangian can be expressed as
\begin{equation}
L_{F_1}=\frac{i}2(\bar\chi_i\tfrac D{dt}\chi^i-\overline{\tfrac D{dt}\chi^i}\chi_i)w^{-1}e^{-3f}-mi\bar\chi_i\ga^0\chi^ie^{-4f}
\end{equation}
in view of the definition of $\chi_0$.
\el
\section{Quantization of the Lagrangian}\las{4}
We consider the functional
\begin{equation}\lae{4.1}
\begin{aligned}
J&=\al_M^{-1}\int_\Om(\bar R-2\Lam)+\int_\Om\tfrac14 \tr(F_{\mu\lam}F^{\mu\lam})\\
&\hp =\;\;\,-\int_\Om\{\tfrac12 \bar g^{\mu\lam}\F_\mu\F_\lam\chi_0^{-\frac13}+U(\F)\chi_0^{-\frac23}\}\\
&\hp =\;\;+\int_\Om\{-\tfrac12[\tilde\psi_iE^\mu_a\ga^a(D_\mu \psi)^i+\overline{\tilde\psi_iE^\mu_a\ga^a(D_\mu \psi)^i}]-m\bar\psi_i\psi^i\chi_0^{-\frac16}\},
\end{aligned}
\end{equation}
where $\al_M$ is a positive coupling constant, $\Om\su N$ is open such that
\begin{equation}
\Om=I\ti\tilde\Om;
\end{equation}
$I=(a,b)$ is a bounded interval and $\tilde\Om\su\so=\R[3]$ an arbitrary open set of measure one with respect to the standard metric of $\R[3]$.

We use the action principle that, for an arbitrary $\Om$ as above, a solution $(A,\F,\psi,\bar g)$ should be a stationary point of the functional with respect to compact variations. This principle requires no additional surface terms for the functional.

Using \frl{3.1}, \frl{3.2}, and \rl{3.3} and arguing as in \cite[section 3]{cg:qfriedman}, where we observe that now $\tilde\ka=0$, we conclude that the functional is equal to 
\begin{equation}\lae{4.3}
\begin{aligned}
J&=\al_M^{-1}\int_a^b\{-6\abs{\dot f}^2e^{3f}w^{-1}-2\Lam e^{3f}w\}\\
&\hp=+3\int_a^b\{(2\abs{\dot\f_0}^2+\abs{\tfrac D{dt}z}^2)w^{-1}e^f-(\f_0^4+8\f_0^2\abs z^2+\abs z^4)we^{-f}\}\\
&\hp= +\int_a^b\{\tfrac12 w^{-1}\abs{\tfrac D{dt}\F}^2e^f-Uwe^{-f}\}\\
&\hp=+\int_a^b\{\frac{i}2(\bar\chi_i\tfrac D{dt}\chi^i-\overline{\tfrac D{dt}\chi^i}\chi_i)-mi\bar \chi_i\ga^0\chi^iwe^{-f}\}.
\end{aligned}
\end{equation}
Here a dot indicates differentiation with respect to the time $t=x^0$ and the covariant derivatives \cq{$\tfrac D{dt}$} of the variables $z,\F,\chi$ are defined in \re{3.21}, \re{3.23}, \fre{3.30}, and in \fre{3.58}.

Thus, our functional depends on the variables $(f,\f_0,z^i,\F^i,\chi^i,w,\Lam^i_j)$. For the variables $w$ and $\Lam^i_j$ no time derivatives exist, i.e., the Legendre transformation will be singular resulting in corresponding constraints. In case of $w$ we obtain the well-known Hamiltonian constraint, while in case of the $\Lam^i_j$ the constraint equations are a bit more complicated. We shall address this issue later.

The dynamical variables are $(f,\f_0,z^i,\F^i,\chi^i)$, where $z^i,\F^i$ are complex and $\chi^i_a$ are anticommuting Grassmann variables. Therefore, we assume that the bosonic and fermionic variables are elements of a graded Grassmann algebra with involution, where the bosonic variables are \tit{even} and the fermionic variables are \tit{odd}. The involution corresponds to the complex conjugation and will be denoted by a bar.

The $\chi^i_a$ are complex variables and we define its real \resp imaginary parts as
\begin{equation}
\xi^i_a=\tfrac1{\sqrt 2}(\chi^i_a+\bar\chi^i_a)
\end{equation}
\resp
\begin{equation}
\h^i_a=\tfrac1{\sqrt 2 i}(\chi^i_a-\bar\chi^i_a).
\end{equation}
Then,
\begin{equation}\lae{4.6.1}
\chi^i_a=\tfrac1{\sqrt 2}(\xi^i_a+i\h^i_a)
\end{equation}
and
\begin{equation}\lae{4.7.1}
\bar\chi^i_a=\tfrac1{\sqrt 2}(\xi^i_a-i\h^i_a).
\end{equation}

In case of even variables we use the usual definitions
\begin{equation}
z^i=x^i+iy^i.
\end{equation}

With these definitions we obtain
\begin{equation}\lae{4.9}
\frac{i}2(\bar\chi_i\tfrac D{dt}\chi^i-\overline{\tfrac D{dt}\chi^i}\chi_i)=\frac{i}2(\xi^a_i\tfrac D{dt}\xi^i_a+\h^a_i\tfrac D{dt}\h^i_a).
\end{equation}

Casalbuoni  quantized a Bose-Fermi system in \cite[section 4]{casalbuoni:fermi} the results of which can be applied to spin $\frac12$ fermions. The Lagrangian in \cite{casalbuoni:fermi} is the same as our Lagrangian in \re{4.9}, and the left derivative is used in that paper, hence we are using left derivatives as well such that  the conjugate momenta of the odd variables are, e.g.,
\begin{equation}
\pi^a_i=\frac{\pa L}{\pa \tfrac D{dt}\xi^i_a}=-\frac i2\xi^a_i,
\end{equation}
and thus the conclusions in \cite{casalbuoni:fermi} can be applied. 

The Lagrangian has been expressed in real variables---at least the important part of it---and it follows that the odd variables $\xi^i_a,\h^i_a$ satisfy, after introducing anticommutative Dirac brackets as in \cite[equ. (4.11)]{casalbuoni:fermi}, 
\begin{equation}
\{\xi^i_a,\xi^j_b\}^*_+=-i\de^{ij}\de_{ab},
\end{equation}
\begin{equation}
\{\h^i_a,\h^j_b\}^*_+=-i\de^{ij}\de_{ab},
\end{equation}
and
\begin{equation}
\{\xi^i_a,\h^j_b\}^*_+=0,
\end{equation}
\cf \cite[equ. (4.19)]{casalbuoni:fermi}.

In view of \re{4.6.1}, \re{4.7.1} we then derive
\begin{equation}
\{\chi^i_a,\bar\chi^j_b\}^*_+=-i\de^{ij}\de_{ab}.
\end{equation}

Canonical quantization---with $\bar h=1$---then requires that the corresponding operators $\hat\chi^i_a, \hat{\bar\chi}^j_b$ satisfy the anticommutative rules
\begin{equation}\lae{4.15.1}
[\hat\chi^i_a,\hat{\bar\chi}^j_b]_+=i\{\chi^i_a,\bar\chi^j_b\}^*_+=\de^{ij}\de_{ab}
\end{equation}
and
\begin{equation}
[\hat{\bar\chi}^i_a,\hat{\bar\chi}^j_b]_+=[\hat\chi^i_a,\hat{\chi}^j_b]_+=0,
\end{equation}
\cf \cite[equ. (3.10)]{casalbuoni:odd} and \cite[equ. (4.17)]{casalbuoni:fermi}.

We could then define a finite dimensional Hilbert space, using Berezin integration, where these
operators would be acting, this is done e.g., in \cite[p.1494]{thiemann:kinematical}, or we could observe, writing $\chi^k_b$ for $\hat\chi^k_b$, etc, that $\chi^k_b$ \resp $\bar\chi^j_c$ can be looked at as being annihilation \resp creation operators in the antisymmetric Fock space, \cf \cite[chap. 65]{dirac:book}; note that Dirac used the reversed symbols for the annihilation and creation operators. 

We adopt the view to represent the operators as operators in the antisymmetric Fock space. Let $\h_0$ be the vacuum vector, normalized to $\norm{\h_0}=1$, then the vector space, where the operators are acting, is spanned by $\h_0$ and by
\begin{equation}\lae{4.13}
\bar\chi^i_{a_1} \bar\chi^i_{a_2}\cdots \bar\chi^i_{a_s}\h_0,
\end{equation}
\begin{equation}\lae{4.14}
\bar\chi^{i_1}_a \bar\chi^{i_2}_a\cdots \bar\chi^{i_r}_a\h_0,
\end{equation}
and mixed products
\begin{equation}\lae{4.15}
\bar\chi^{i_1}_a \bar\chi^{i_2}_a\cdots \bar\chi^{i_r}_b \bar\chi^{i_r}_c\h_0,
\end{equation}
where all operators acting on $\h_0$ have to be different otherwise the result will vanish. Hence, the vector space is a finite dimensional subspace of the antisymmetric Fock space.

Defining the number operator
\begin{equation}
n^i_a= \bar\chi^i_a\chi^i_a,
\end{equation}
we deduce from \re{4.15.1}
\begin{equation}
\chi^i_a \bar\chi^i_a=I-n^i_a.
\end{equation}

The vacuum vector $\h_0$ belongs to the kernel of all $n^i_a$, hence we have
\begin{equation}
\chi^i_a \bar\chi^i_a\h_0=\h_0.
\end{equation}
$\chi^i_a$ and $\bar\chi^i_a$ are adjoints of each other, i.e., $n^i_a$ is self-adjoint, and there holds
\begin{equation}
\chi^i_a\h_0=0\qq\qq\A\,(a,i)
\end{equation}
in view of
\begin{equation}
0=n^i_a\h_0= \bar\chi^i_a\chi^i_a\h_0.
\end{equation}
Moreover, the vectors in \re{4.13}, \re{4.14} and \re{4.15} are normalized eigenvectors of $n^i_a$ with eigenvalues $1$ \resp $0$ depending on the fact if $\bar\chi^i_a$ happens to be acting on $\h_0$ or not.

The fermionic Hamiltonian is equal to
\begin{equation}
H_{F_1}=mi\bar\chi_i\ga^0\chi^iwe^{-f}.
\end{equation}
Using the definition of $\ga^0$,
\begin{equation}
\ga^0=i\begin{pmatrix}
\I&0\\[\cma]
0&-\I
\end{pmatrix},
\end{equation}
we deduce
\begin{equation}
i\bar\chi_i\ga^0\chi^i=-(\bar\chi^{\bar a}_i\chi^i_{\bar a}-\bar\chi^{\underline a}_i\chi^i_{\underline a}),
\end{equation}
where
\begin{equation}
1\le\bar a\le 2\q\wed\q 3\le \underline a\le 4
\end{equation}
with similar definitions for $\bar b$, $\underline b$, etc.

Hence, we conclude
\begin{equation}
H_{F_1}=m(\bar\chi^{\underline a}_i\chi^i_{\underline a}-\bar\chi^{\bar a}_i\chi^i_{\bar a}) w e^{-f}
\end{equation}
where of course the factor $we^{-f}$ will be taken care of when we shall consider the full Hamiltonian and the Hamiltonian constraint \resp the Wheeler-DeWitt equation. Note that the sign of $m$ is irrelevant for our considerations. However, for definiteness, we shall assume $m>0$.

Let us now quantize the bosonic part. Without changing the notation we shall assume that the complex fields $\F$, $z$ have real valued components by doubling their dimensions, i.e., $\F$ and $\z$ now have $2n$ real components
\begin{equation}
\F=(\F^i)\q\wed\q z=(z^i),\q 1\le i\le 2n.
\end{equation}

Before we apply the Legendre transformation, let us express the quadratic derivative terms with the help of a common metric.

For $0\le A,B\le 4n+1$, define
\begin{equation}
(y^A)=(f,\f_0,z^i,\F^i),
\end{equation}
\begin{equation}
(G_{AB})=\diag(-12\al_M^{-1} e^{2f},12,6\I_{2n},\I_{2n})e^f,
\end{equation}
and
\begin{equation}
V=3(\f_0^4+8\f_0^2\abs z^2+\abs z^4).
\end{equation}
Then $J$ in \re{4.3} can be expressed as
\begin{equation}\lae{4.32}
\begin{aligned}
J&=\int_a^bw\{G_{AB}\tfrac D{dt}y^A\tfrac D{dt}y^Bw^{-2}-2\al_M^{-1}\Lam e^{3f}-Ve^{-f}-Ue^{-f}\} \\
&\hp=+\int_a^b\{\frac{i}2(\bar\chi_i^a\tfrac D{dt}\chi^i_a-\overline{\tfrac D{dt}\chi^i_a}\chi_i^a)-m(\bar\chi^{\underline a}_i\chi^i_{\underline a}-\bar\chi^{\bar a}_i\chi^i_{\bar a})e^{-f}w\}.
\end{aligned}
\end{equation}

Applying now the Legendre transformation we obtain the Hamiltonian 
\begin{equation}
\begin{aligned}
\tilde H&=\tilde H(w,y^A,p_A,\xi^i_a,\h_a^i)=p_A\tfrac D{dt}y^A+\tfrac D{dt}\xi^i_a \pi^a_i +\tfrac D{dt}\h^i_a\s^a_i -L\\
&= \{\tfrac12 G_{AB}\tfrac D{dt}y^A\tfrac D{dt}y^Bw^{-2}+2\al_M^{-1}\Lam e^{3f}+Ve^{-f}+Ue^{-f}\}w\\
&\hp=\q+m(\bar\chi^{\underline a}_i\chi^i_{\underline a}-\bar\chi^{\bar a}_i\chi^i_{\bar a})e^{-f}w\\
&= \{\tfrac12 G^{AB}p_Ap_B+2\al_M^{-1}\Lam e^{3f}+Ve^{-f}+Ue^{-f}\}w\\
&\hp=\q+m(\bar\chi^{\underline a}_i\chi^i_{\underline a}-\bar\chi^{\bar a}_i\chi^i_{\bar a})e^{-f}w\\
&\equiv Hw,
\end{aligned}
\end{equation}
and the Hamiltonian constraint requires 
\begin{equation}
H(y^A,\chi^i_a,\bar\chi^i_a,p_A)=0.
\end{equation}

Canonical quantization stipulates that, in case of the bosonic variables, we replace the momenta $p_A$ by
\begin{equation}
p_A=-i\frac\pa{\pa y^A},
\end{equation}
where $\hbar=1$, and for the fermionic variables we consider $\bar\chi^i_a$ and $\chi^a_i$ as creation \resp annihilation operators  in a $2^{4n}$ dimensional subspace $\mc F_1$ of the antisymmetric Fock space as described above                                     .

Thus, the Hamilton operator is equal to
\begin{equation}\lae{4.36}
\begin{aligned}
H=-\tfrac12 \D+(V+U)e^{-f}+2\al_M^{-1}\Lam e^{3f}+m(\bar\chi^{\underline a}_i\chi^i_{\underline a}-\bar\chi^{\bar a}_i\chi^i_{\bar a})e^{-f},
\end{aligned}
\end{equation}
where the metric $G_{AB}$ is a Lorentz metric, i.e., the bosonic part of $H$ is hyperbolic.

Ignoring for the moment a crucial first-class constraint we haven't considered yet, which is due to the variables $\Lam^i_j$, we have to find wave functions
\begin{equation}
\Psi=\Psi(y),
\end{equation}
where
\begin{equation}
\Psi:\R[4n+2]\ra \mc F_1,
\end{equation}
such that
\begin{equation}
H\Psi=0;
\end{equation}
moreover, we even have to find a spectral resolution of this problem.

We shall consider wave functions of the form
\begin{equation}
\Psi(y)=u(y)\otimes\h,\q \h\in\mc F_1,
\end{equation}
where $u$ belongs to a suitable Hilbert space consisting of complex valued functions.

Let $\Psi=u\otimes \h$ be a smooth functions, then
\begin{equation}
\D\Psi=\tfrac1{\sqrt{\abs G}}\frac\pa{\pa y^A}(\sqrt{\abs G}G^{AB}\Psi_B).
\end{equation}

Now,
\begin{equation}
\abs G=864 \al_M^{-1}e^{4(n+1)f},
\end{equation}
and hence
\begin{equation}
-\D\Psi=\tfrac1{12} e^{-2(n+1)f}\frac\pa{\pa y^0}(e^{(2n-1)f}\frac{\pa\Psi}{\pa y^0})-2a^{\al\bet}\Psi_{\al\bet}e^{-f}-\tilde\D\Psi e^{-f}, 
\end{equation}
where $(a^{\al\bet})$ is a positive definite diagonal matrix
\begin{equation}
(a^{\al\bet})=\diag(\tfrac1{24},\tfrac1{12}\I_{2n}),
\end{equation}
and the indices range from $1\le\al,\bet\le 2n+1$, and $\tilde\D$ is the Laplacian with respect to the $2n$ variables $\F^i$. $\Psi_{\al\bet}$ are ordinary partial derivatives of $\Psi$.

Thus, we deduce from \re{4.36} that the Wheeler-DeWitt equation looks like
\begin{equation}
\begin{aligned}
&\tfrac1{24} e^{-2(n+1)f}\frac\pa{\pa y^0}(e^{(2n-1)f}\frac{\pa\Psi}{\pa y^0})-a^{\al\bet}\Psi_{\al\bet}e^{-f}-\tilde \D\Psi e^{-f}\\
&\q+(V+U)\Psi e^{-f}+2\al_M^{-1}\Lam e^{3f}\Psi+m(\bar\chi^{\underline a}_i\chi^i_{\underline a}-\bar\chi^{\bar a}_i\chi^i_{\bar a})\Psi e^{-f}=0.
\end{aligned}
\end{equation}

Multiplying this equation by $e^f$ we have proved:
\bt
The Wheeler-DeWitt equation for the functional $J$ in \re{4.3} has the form
\begin{equation}
H_1\Psi+H_2\Psi+H_{F_1}\Psi-H_0\Psi=0,
\end{equation}
where
\begin{equation}
H_0\Psi=-\tfrac1{24} e^{-(2n+1)f}\frac\pa{\pa y^0}(e^{(2n-1)f}\frac{\pa\Psi}{\pa y^0})-2\al_M^{-1}\Lam e^{4f}\Psi,
\end{equation}
\begin{equation}
H_1\Psi=-a^{\al\bet}\Psi_{a\bet}+V\Psi,
\end{equation}
\begin{equation}
H_2\Psi=-\tfrac12\tilde\D\Psi+U\Psi,
\end{equation}
and
\begin{equation}
H_{F_1}\Psi=m(\bar\chi^{\underline a}_i\chi^i_{\underline a}-\bar\chi^{\bar a}_i\chi^i_{\bar a})\Psi.
\end{equation}
\et

We emphasize that $y^0$ and $f$ denote the same real variable.

Before we can solve the Wheeler-DeWitt equation we still have to formulate and satisfy the first-class constraint resulting from the presence of the variables $\Lam^i_j$. This will be done in the next section.

\section{A first-class constraint}\las{5}

The Lagrangian functional in the previous section contains as non-dynamical variables the $\Lam^i_j$, besides the $w$, which has already been taken care of by the Hamiltonian constraint.

The requirement that the first variation of the functional  with respect to compact variations of \tit{all} variables should vanish leads to a set of constraint equations due to the presence of the $\Lam^i_j$.

$(\Lam^i_j)$ is an arbitrary antisymmetric matrix in $\Ccc$ with trace zero if $n>1$.

To compute the first variation of $J$ with respect to the $\Lam^i_j$, we look at the integral in \fre{4.32}. Since we also have to differentiate the Dirac term it is best to rewrite the quadratic form
\begin{equation}
\tfrac12 G_{AB}\tfrac D{dt}y^A\tfrac D{dt}y^Bw^{-1} 
\end{equation}
in the form
\begin{equation}
\tfrac12 G_{AB}\tfrac D{dt}y^A\overline{\tfrac D{dt}y^B}w^{-1},
\end{equation}
where
\begin{equation}
(y^A)=(f,\f_0,z^i,\zeta^i);
\end{equation}
$z^i$, $\zeta^i$ are complex components and $\zeta^i$ symbolizes $\F^i$.

The terms involved are
\begin{equation}\lae{5.4}
\begin{aligned}
\tfrac12 G_{AB}\tfrac D{dt}y^A\overline{\tfrac D{dt}y^B}w^{-1} +\frac{i}2(\bar\chi_i\tfrac D{dt}\chi^i-\overline{\tfrac D{dt}\chi^i}\chi_i).
\end{aligned}
\end{equation}

Let us first look at the bosonic term and because of the symmetry it suffices to consider the $z^i$.

The independent components of $(\Lam^i_j)$ can be labelled as
\begin{equation}
\Lam^k_m,\qq 1\le k<m\le n,
\end{equation}
and
\begin{equation}
\Lam^k_k, 1\le k\le n-1,
\end{equation}
if $n>1$, no summation over  $k$. Since $\tr(\Lam^i_j)=0$, we assume the first $(n-1)$ diagonal elements to be independent imaginary variables and
\begin{equation}
\Lam^n_n=-\sum_{k=1}^{n-1}\Lam^k_k.
\end{equation}

Let us start with a component
\begin{equation}\lae{5.8}
\Lam^k_m=a+ib
\end{equation}
for $1\le k<m\le n$.

By observing that
\begin{equation}
p_A=G_{AB}\tfrac D{dt}y^Bw^{-1},
\end{equation}
we deduce that the terms in \re{5.4} involving the numbers $a,b$ are
\begin{equation}
\begin{aligned}
\tfrac12\{p_k\bar\Lam^k_m\bar z^m+p_m\bar\Lam^m_k\bar z^k+\bar p_k\Lam^k_mz^m+\bar p_m\Lam^m_kz^k\},
\end{aligned}
\end{equation}
or equivalently,
\begin{equation}
\begin{aligned}
\tfrac12\{p_k(a-ib)\bar z^m-p_m(a+ib)\bar z^k+\bar p_k(a+ib)z^m-\bar p_m(a-ib)z^k\}.
\end{aligned}
\end{equation}

Differentiating first with respect to $\frac\pa{\pa a}$ we obtain
\begin{equation}\lae{5.12}
\tfrac12\{\bar p_kz^m-p_m\bar z^k\}+\tfrac12\{-\bar p_mz^k+p_k\bar z^m\},
\end{equation}
and differentiating with respect to $-i\frac\pa{\pa b}$ yields
\begin{equation}\lae{5.13}
\tfrac12\{\bar p_kz^m-p_m\bar z^k\}-\tfrac12\{-\bar p_mz^k+p_k\bar z^m\}.
\end{equation}

Differentiating the diagonal terms we obtain
\begin{equation}\lae{5.14}
\tfrac12\{\bar p_kz^k-p_k\bar z^k\}-\tfrac12\{\bar p_nz^n-p_n\bar z^n\}
\end{equation}
for $1\le k\le n-1$, and
\begin{equation}\lae{5.15}
\tfrac12\{\bar pz-p\bar z\}\tfrac43
\end{equation}
in case $n=1$.

Looking at the terms in \re{5.12} and \re{5.13} we see they represent the real \resp imaginary part of the complex term
\begin{equation}\lae{5.16}
\bar p_k z^m-p_m\bar z^k,\qq 1\le k<m\le n.
\end{equation}     
Note that the variables are still complex Grassmann variables and not yet operators. 

When formulating the constraint equations, the terms in \re{5.12}, \re{5.13} will be set to vanish. Hence, these equations are equivalent to the complex equations
\begin{equation}\lae{5.17}
\bar p_k z^m-p_m\bar z^k=0,\qq 1\le k<m\le n,
\end{equation}     
as well as to their complex conjugates
\begin{equation}\lae{5.18}
p_k \bar z^m-\bar p_m z^k=0,\qq 1\le k<m\le n.
\end{equation}     
\br\lar{5.1}
 After quantization the left-hand sides of the equations above will be linear operators in a space of complex valued test functions. It will turn out that the operator  resulting from \re{5.18} will be the adjoint of the operator resulting from \re{5.17}, what is already evident since the quantization process will turn complex conjugation into forming the adjoint. 
 \er
Similar arguments apply when we differentiate the Dirac terms. The terms in \re{5.12} \resp \re{5.13} will then correspond to
\begin{equation}
ig_1\{\bar\chi^a_k\chi^m_a-\bar\chi^a_m\chi^k_a\}
\end{equation}
\resp
\begin{equation}
ig_1\{\bar\chi^a_k\chi^m_a+\bar\chi^a_m\chi^k_a\},
\end{equation}
hence, the equivalent to \re{5.17} will be 
\begin{equation}\lae{5.20}
2ig_1\bar\chi^a_k\chi^m_a,
\end{equation}
and  the equivalent of \re{5.18}
\begin{equation}\lae{5.22d}
-2ig_1\bar \chi^m_a \chi^a_k.
\end{equation}

The diagonal term has the form
\begin{equation}\lae{5.21}
ig_1\{\bar\chi^a_k\chi^k_a-\bar\chi^a_n\chi^n_a\},\qq 1\le k<n,
\end{equation}
where the summation convention is not used for the index $k$, but of course for the index $a$.
In case $n=1$ we have
\begin{equation}\lae{5.22}
ig_1\bar\chi^a\chi_a.
\end{equation}

Since we shall later, after quantization, when these terms have turned into operators, apply the operators to complex valued wave functions, we consider the complex expressions as the primary terms to determine the constraints. 

The full  constraint equations are 
\begin{equation}\lae{5.25d}
l_{k,m}+g_1\tilde l_{k,m}+g_1\hat l_{k,m}=0,\qq 1\le k<m\le n,
\end{equation}
or equivalently, their complex conjugates, 
\begin{equation}\lae{5.26d} 
\bar l_{k,m}+g_1\bar{\tilde l}_{k,m}+g_1\bar{\hat l}_{k,m}=0,\qq 1\le k<m\le n,
\end{equation}
\begin{equation}
l_k+g_1\tilde l_k+g_1\hat l_k=0,\qq 1\le k<n,
\end{equation}
and
\begin{equation}
l_0+g_1\tilde l_0+g_1\hat l_0=0,\qq n=1,
\end{equation}
where $l_{k,m}$,  $l_k$ \resp $l_0$ represent the terms in \re{5.17}, \re{5.14} \resp \re{5.15}, $\hat l_{k,m}$, $\hat l_k$, \resp $\hat l_0$ are defined by the equations  \re{5.20}, \re{5.21} \resp \re{5.22}, while
\begin{equation}
\tilde l_{k,m}=\{\bar\pi_k\zeta^m-\pi_m\bar\zeta^k\},
\end{equation}
\begin{equation}
\tilde l_k=\tfrac12\{\bar\pi_k\zeta^k-\pi_k\bar\zeta^k\}-\tfrac12\{\bar\pi_n\zeta^n-\pi_n\bar\zeta^n\},
\end{equation}
and
\begin{equation}
\tilde l_0=\tfrac12\{\bar\pi \zeta-\pi\bar\zeta\}\tfrac43.
\end{equation}

The coupling constant $g_1$ appears because it entered into the definition of the covariant derivatives of $\F$ and $\chi$, but not in the case of $z$.

The constraint equations are first-class constraints, according to Dirac, after quantization they have to be satisfied by the wave functions.

The terms for the fermionic variables can already be looked at as operators in the antisymmetric Fock space. For the  quantization of the bosonic terms, we only consider $l_{k,m}$, $l_k$ and $l_0$. Writing
\begin{equation}
p_k=p_{x^k}+ip_{y^k}
\end{equation}
and
\begin{equation}
\bar p_k=p_{x^k}-ip_{y^k}
\end{equation}
and replacing $p_k$, $\bar p_k$ by the operators
\begin{equation}
p_k\q\ra\q -i\{\tfrac\pa{\pa x^k}+i\tfrac\pa{\pa y^k}\},
\end{equation}
\begin{equation}
\bar p_k\q\ra\q -i\{\tfrac\pa{\pa x^k}-i\tfrac\pa{\pa y^k}\}
\end{equation}
we deduce from \re{5.17}, \re{5.14}, and \re{5.15}, without changing the notation,
\begin{equation}\lae{5.34}
\begin{aligned}
l_{k,m}&=\big(y^k\frac\pa{\pa x^m} -x^m\pde{}{y^k}\big)+\big(y^m\pde{}{x^k}-x^k\pde{}{y^m}\big) \\
&\hp=+i\Big\{\big(x^k\pde{}{x^m}-x^m\pde{}{x^k}\big)+\big(y^k\pde{}{y^m}-y^m\pde{}{y^k}\big)\Big\},
\end{aligned} 
\end{equation}
\begin{equation}
l_k=\big(x^k\pde{}{y^k}-y^k\pde{}{x^k}\big)-\big(x^n\pde{}{y^n}-y^n\pde{}{y^n}\big),
\end{equation}
and 
\begin{equation}\lae{5.35}
l_0=\tfrac43\big(x\pde{}y-y\pde{}x\big)+\tfrac83i.
\end{equation}

When we use the formulation \re{5.18} instead of  \re{5.17} the operator $l_{k,m}$ in \re{5.34} will be replaced by its formal adjoint
\begin{equation}\lae{5.37}
\begin{aligned}
l^*_{k,m}&=-\big(y^k\frac\pa{\pa x^m} -x^m\pde{}{y^k}\big)-\big(y^m\pde{}{x^k}-x^k\pde{}{y^m}\big) \\
&\hp=+i\Big\{\big(x^k\pde{}{x^m}-x^m\pde{}{x^k}\big)+\big(y^k\pde{}{y^m}-y^m\pde{}{y^k}\big)\Big\}.
\end{aligned} 
\end{equation}

The differential operators $\tilde l_{k,m}$, etc., are similarly defined; we shall denote the corresponding variables by $\tilde x^i$ and $\tilde y^i$, $1\le i\le n$.

To solve the Wheeler-DeWitt equation we have to define a Hilbert space generated by wave functions $\Psi$ satisfying the constraint equations
\begin{equation}\lae{5.40}
(l_{k,m}+g_1\tilde l_{k,m}+g_1\hat l_{k,m})\Psi=0,
\end{equation}
or equivalently,
\begin{equation}\lae{5.41}
(l^*_{k,m}+g_1\tilde l^*_{k,m}+g_1\hat l^*_{k,m})\Psi=0,
\end{equation}
and
\begin{equation}
(l_k+g_1\tilde l_k+ g_1\hat l_k)\Psi=0.
\end{equation}
In case $n=1$,
\begin{equation}
(l_0+g_1\tilde l_0+g_1\hat l_0)\Psi=0.
\end{equation}

 Later we shall define various Hilbert spaces and before defining a Hilbert space we shall deliberately decide which constraint formulation, either \re{5.25d} or \re{5.26d},  we shall use at the classical level, where both formulations are equivalent, since it will make an important difference after quantization.

The Hilbert spaces will be tensor products, where, to address the constraint equations, it suffices to restrict our attention to wave functions of the form
\begin{equation}
\Psi=u(z,\tilde z)\otimes \h,
\end{equation}
where $(z,\tilde z)\in \R[4n]=\R[2n]\times\R[2n]$ and $\h$ belongs to the antisymmetric Fock space. Occasionally, we also use the symbol $\zeta$ instead of $\tilde z$.

To solve the constraint equations, we consider each factor $u$ and $\h$ separately. 

$\h$ belongs to a finite dimensional subspace $\mc F_1$. Define the linear map
\begin{equation}
\lam_0=(\hat l_{k,m})_{1\le k<m\le n}:\mc F_1\ra \mc F_1^{\frac{n(n-1)}2}.
\end{equation}

Let $\hat{\mc F}_0$ be the image of
\begin{equation}
\mc F_1\ni\h \ra \h\equiv (\h,\ldots,\h)\in \mc F_1^{n-1}, 
\end{equation}
and $\Lam_0$ be the map
\begin{equation}
\Lam_0=(\hat l_k)_{1\le k<n}:\hat{\mc F}_0\ra \mc F_1^{n-1}.
\end{equation}

We then look for eigenspaces  of $-i\Lam_0$
\begin{equation}
\tilde F_{\s}=\set{\h\in \hat{\mc F}_0}{-i\Lam_0\h=\s\h},
\end{equation}
where we identify $\h$ and $(\h,\ldots,\h)$, i.e., we especially consider 
\begin{equation}\lae{5.45}
\tilde F_\s\su\mc F_1.
\end{equation}
\bl\lal{5.1b}
The eigenvalues $\s$ of $-i\Lam_0$ belong to the set
\begin{equation}
M_4=\{-4,-3,\ldots,0,\ldots, 3,4\}
\end{equation}
and each possible eigenvalue is assumed. The $\tilde F_\s$ are mutually orthogonal. 
\el

\bp
(i) The claim that the eigenvalues are elements of $M_4$ will be proved in \rl{5.3a}.

(ii) In order to prove that every element of $M_4$ is indeed an eigenvalue we shall give a list of eigenvectors belonging to $\tilde F_\s$ for each $\s\in M_4$.
\begin{equation}
\bar\chi^n_1\cdots \bar\chi^n_4\h_0\in \tilde F_{-4},
\end{equation}
\begin{equation}
\bar\chi^n_1\bar\chi^n_2\bar\chi^n_3\h_0\in \tilde F_{-3},
\end{equation}
\begin{equation}
\bar\chi^n_1\bar\chi^n_2\h_0\in \tilde F_{-2},
\end{equation}
\begin{equation}
\bar\chi^n_b\h_0\in \tilde F_{-1},
\end{equation}
\begin{equation}
\h_0\in \tilde F_0.
\end{equation}
For $1\le b\le 4$ define
\begin{equation}
\h_b=\bar\chi^1_1\cdots\bar\chi^1_b\cdots\bar\chi^{n-1}_1\cdots\bar\chi^{n-1}_b\h_0,
\end{equation}
then
\begin{equation}
\h_b\in \tilde F_{b}.
\end{equation}

Since the eigenvectors are especially eigenvectors of the self-adjoint operator $-i\hat l_1$, eigenvectors belonging to different eigenvalues are orthogonal.
\ep

\bl\lal{5.2a}
Let $\bar\chi^a_k$, $\chi^k_a$, $1\le a\le m_1$, where $k$ is fixed, be creation \resp annihilation operators in the antisymmetric Fock space, then the eigenvalues of 
\begin{equation}
l_k=\bar\chi^a_k\chi^k_a,
\end{equation}
where we use summation over $a$, belong to the set
\begin{equation}
M_1=\{0,1,\ldots,m_1\}.
\end{equation}
\el
\bp
We use induction with respect to $m_1$. When $m_1=1$ this result is due to the fact that a number operator is a projector. 

Thus assume that the claim has already been proved for $m_1< m$ with  $m>1$ and set $m_1=m$. Let $\lam$ be an eigenvalue of $l_k$ and $\h$ an eigenvector. Then we write $\h$ as
\begin{equation}\lae{5.56a}
\h=\h_1+\h_2,
\end{equation}
where $\h_1$ can be written in the form
\begin{equation}
\h_1=\bar\chi^1_k\xi
\end{equation}
and $\h_2$ can be written as a linear combination of standard basis vectors which do not contain the creation operator $\bar\chi^1_k$. Hence, $\h_2$ belongs to the kernel of $\bar\chi^1_k\chi^k_1$ and we deduce
\begin{equation}\lae{5.58a}
\lam \h_1+\lam \h_2=l_k\h=\h_1+\sum_{a=2}^{m}\bar\chi^a_k\chi^k_a\h.
\end{equation}
Let $\bar\chi^1_k$ act on both sides of this equation then
\begin{equation}
\lam\bar\chi^1_k\h_2=\sum_{a=2}^{m}\bar\chi^a_k\chi^k_a\bar\chi^1_k\h_2
\end{equation}
and we conclude either that $0\le \lam\le m-1$ or that $\h_2=0$.

Suppose $\h_2=0$, then, in view of \re{5.58a}, we obtain
\begin{equation}
(\lam -1)\h_1=\sum_{a=2}^{m}\bar\chi^a_k\chi^k_a\h_1
\end{equation}
yielding
\begin{equation}
0\le \lam\le m
\end{equation}
because of the induction hypothesis. 
\ep
\bl\lal{5.3a}           
Let $\bar\chi^a_k$, $\chi^k_a$, $\bar\chi^b_n$, $\chi^n_b$, $1\le a\le m_1$, $1\le b\le m_2$, where $k,n$, $k\ne n$,  are fixed, be creation \resp annihilation operators in the antisymmetric Fock space, then the eigenvalues of 
\begin{equation}
l=\bar\chi^a_k\chi^k_a-\bar\chi^b_n\chi^n_b,
\end{equation}
where we use summation over $a$ and $b$, belong to the set
\begin{equation}
M_1=\{-m_2, -m_2+1,\dots, 0,1,\ldots,m_1\}.
\end{equation}
\el
\bp
We use induction with respect to $m_2$. Actually we only prove it for $m_2=1$ and refer for the further steps in the induction arguments to the proof of the preceding lemma. Thus, let $m_2=1$ and let $\lam$ be an eigenvalue of $l$ with eigenvector $\h$. Split $\h$ similarly as in \re{5.56a}
\begin{equation}
\h=\h_1+\h_2,
\end{equation}
where now 
\begin{equation}
\h_1=\bar\chi^1_n\xi.
\end{equation}
Then, we infer
\begin{equation}\lae{5.66a}
\lam\h_1+\lam\h_2=l\h=l_k\h-\h_1,
\end{equation}
and conclude further, as in the proof before,
\begin{equation}
l_k\bar\chi^1_n\h_2=\lam\bar\chi^1_n\h_2,
\end{equation}
hence, we either have $0\le \lam\le m_1$, in view of \rl{5.2a}, or $\h_2=0$. The latter would imply, because of \re{5.66a}, 
\begin{equation}
l_k\h_1=(\lam+1)\h_1,
\end{equation}
completing the proof of the lemma.
\ep
\bd
Let $\tilde F_{\s_i}$ be one of the eigenspaces in \rl{5.1b}, then we define in case $\s_i\ge 0$
\begin{equation}
F_{\s_i}=\set{\h\in \tilde F_{\s_i}}{\hat l_{k,m}\h=0\q\A\, 1\le k<m\le n}
\end{equation}
and in case $\s_i< 0$
\begin{equation}
F_{\s_i}=\set{\h\in \tilde F_{\s_i}}{\hat l^*_{k,m}\h=0\q\A\, 1\le k<m\le n}.
\end{equation}
\ed
\br
The fermions defined in \rl{5.1b} which belong to $\tilde F_{\s_i}$ also belong to $F_{\s_i}$. Hence, we have  
\begin{equation}
\dim F_{\s_i}\ge 1\qq\A\, 1\le i\le 9.
\end{equation}
\er
The eigenspace $F_0$, i.e., $\s_i=0$,   will be of special importance, since it contains the $\SU(3)$ fermions used in forming the quarks, when $n=3$, as we shall prove:  
\bl\lal{5.7b}
Let $n\ge 2$, then the dimension of the eigenspace $F_0$  is at least 16.  It contains the mutually orthogonal unit vectors
\begin{equation}
\bar \chi^1_M\cdots \bar \chi^n_M\h_0\qq\A\,M\in \mc P(\{1,2,3,4\}),
\end{equation}
where $\mc P(\{1,2,3,4\})$ is the power set of $\{1,2,3,4\}$, and the operators $\bar\chi^k_M$ are defined by
\begin{equation}
\bar\chi^k_M=
\begin{cases}
\I, &M=\eS,\\
\bar\chi^k_{a_1}\cdots \bar\chi^k_{a_i},&M=\{a_1,\ldots,a_i\},
\end{cases}
\end{equation}
where, for definiteness, the factors in the  product are ordered by the standard order of the natural numbers, i.e., in the definition above, we assume
\begin{equation}
a_1<a_2<\cdots<a_i.
\end{equation}
\el
\bp
Easy exercise.
\ep

Next, we fix an eigenvalue $\s_i$ with corresponding eigenspace $F_{\s_i}$, where we emphasize the convention \re{5.45}, and we want to define a matching bosonic Hilbert space $\mc H(\s_i)$ such that
\begin{equation}
l_qu=0\q\wed\q \tilde l_qu=-i\s_iu\qq\A\, u\in\mc H(\s_i),
\end{equation}
and $1\le q<n$, and such that
\begin{equation}
l_{k,m}u=0\q\wed\q \tilde l_{k,m}u=0\qq\A\, u\in\mc H(\s_i),
\end{equation}
for all $1\le k<m\le n$, if $\s_{i}\ge 0$, and 
\begin{equation}
l^*_{k,m}u=0\q\wed\q \tilde l^*_{k,m}u=0\qq\A\, u\in\mc H(\s_i),
\end{equation}
for all $1\le k<m\le n$, if $\s_i<0$.

\br\lar{5.2}
The Hilbert spaces
\begin{equation}
\mc H(\s_i)\otimes F_{\s_i}
\end{equation}
would then be mutually orthogonal and its elements would satisfy the constraints.
\er

We shall show that this procedure is always possible; we formulate and prove the result for  generic differential operators $l_{k,m},l_k$, \resp for $l^*_{k,m}$, $l_k$, and for $n\ge 2$---the case $n=1$ will be dealt with in \rs{8}.  
\bt\lat{5.3}
For any $r\in \N$ there exists a largest infinite dimensional subspace
\begin{equation}
E\su C^\un_c(\R[2n],\Cc)
\end{equation}
such that all $u\in E$ satisfy
\begin{equation}
l_{k,m}u=0\qq\A\,1\le k<m\le n
\end{equation}
and
\begin{equation}
l_ku=-iru\qq\A\,1\le k<n.
\end{equation}
Moreover, let $V(z)=V_0(\abs z^2)$ be a smooth potential, $V_0\in C^\un(\R[])$, then $E$ is invariant with respect to the operators
\begin{equation}\lae{5.62}
u\ra Vu
\end{equation}
and
\begin{equation}
u\ra \D u.
\end{equation}
\et
\bp
 We first prove that there exists an infinite dimensional subspace with the above properties. For any $\rho\in C^\un_c(\R[])$ the function
\begin{equation}\lae{5.63}
\f=\rho(\abs z^2)
\end{equation}
satisfies
\begin{equation}
l_{k,m}\f=0\q\wed\q l_k\f=0.
\end{equation}
Let
\begin{equation}
u_n=x^n+iy^n,
\end{equation}
then
\begin{equation}
l_ku_n=-iu_n\qq\A\, 1\le k<n
\end{equation}
and
\begin{equation}
l_{k,m}u_n=0\qq\A\, 1\le k<m\le n.
\end{equation}

Since $l_k,l_{k,m}$ are linear differential operators of first order we infer that
\begin{equation}
u=u_n^r
\end{equation}
satisfies
\begin{equation}
l_ku=-iru\qq\A\, 1\le k<n.
\end{equation}
Let $\rho\in C^\un_c(\R[])$ be arbitrary and define
\begin{equation}\lae{5.71}
v=u\f, \qq\f=\rho(\abs z^2),
\end{equation}
then $v$ is smooth and
\begin{equation}
l_kv=-irv\qq\A\,1\le k<n,
\end{equation}
as well as
\begin{equation}
l_{k,m}v=0.
\end{equation}

Since the support of $\rho$ is arbitrary, the functions $v$ in \re{5.71} generate an infinite dimensional subspace $\tilde E\su C^\un_c(\R[2n],\Cc).$ 

Obviously, $\tilde E$ is invariant with respect to the operator in \re{5.62}. It remains to prove the invariance with respect to the Laplace operator.

An immediately calculation reveals
\begin{equation}
\D u_n^r=0,
\end{equation}
\begin{equation}
\D \f=4n\dot \rho+4\Ddot\rho\abs z^2,
\end{equation}
\begin{equation} \lae{5.77}
D_iu_n^rD^i\f=2ru_n^r\dot\rho,
\end{equation}
and
\begin{equation}
\D(u_n^r\f)=(4n\dot\rho+4\Ddot\rho\abs z^2)u^r_n+4ru_n^r\f.
\end{equation}

Thus, $\tilde E\su C^\un_c(\R[2n],\Cc)$ is infinite dimensional and invariant for $V$ and $\D$, and its elements satisfy the constraint equations. To define a largest subspace with these properties, we consider the family
\begin{equation}
\mc F=\set{F\su C^\un_c(\R[2n],\Cc)}{F\;\tup{subspace with the above properties.}}
\end{equation}
$\mc F\ne \eS$ and the space generated by
\begin{equation}
E=\uuu_{F\in\mc F}F
\end{equation}
is the largest subspace with these properties as one easily checks, and hence $E$ is the largest subspace.
\ep
\bt\lat{5.8}
For any $r\in \N$ there exists a largest infinite dimensional subspace
\begin{equation}
E\su C^\un_c(\R[2n],\Cc)
\end{equation}
such that all $u\in E$ satisfy
\begin{equation}
l^*_{k,m}u=0\qq\A\,1\le k<m\le n
\end{equation}
and
\begin{equation}
l_ku=iru\qq\A\,1\le k<n.
\end{equation}
Moreover, let $V(z)=V_0(\abs z^2)$ be a smooth potential, $V_0\in C^\un(\R[])$, then $E$ is invariant with respect to the operators
\begin{equation}
u\ra Vu
\end{equation}
and
\begin{equation}
u\ra \D u.
\end{equation}
\et
\bp
In view of the proof of the preceding theorem it suffices to show that
\begin{equation}
u_n=x^n-iy^n
\end{equation}
satisfies
\begin{equation}
l_ku_n=iu_n\qq\A\, 1\le k<n
\end{equation}
and
\begin{equation}
l^*_{k,m}u_n=0\qq\A\, 1\le k<m\le n,
\end{equation}
but these equations follow immediately.
\ep

\br\lar{5.4}
In the preceding two theorems the elements of $E$ are eigenfunctions of $l_k$ with integer eigenvalues, which will suffice for our purposes, since the corresponding eigenvectors of the fermionic operators $\hat l_k$ will also have integer eigenvalues. But even in a situation when the possible eigenvalues of the $\hat l_k$ would be multiples of a given positive number $\lam$ we could define a matching bosonic Hilbert space by modifying the definition of the covariant differentiation of the Higgs field. Instead of the definition \fre{3.27} we would then define
\begin{equation}
\F_\mu=\F_{,\mu}+\lam g_1\hat A_\mu\F,
\end{equation}
\er
 
\br\lar{5.5}
If the potential $V$ depends on additional variables $\xi=(\xi^i)$, $1\le i\le m$,
\begin{equation}
V=V_0(\abs z^2,\xi),
\end{equation}
which do not enter into the constraint equations, then a largest subspace can be constructed by choosing the test functions $\f$ in \re{5.63} to be of the form
\begin{equation}
\f=\rho(\abs z^2,\xi),
\end{equation}
with
\begin{equation}
\rho\in C^\un_c(\R[]\times\R[m],\Cc).
\end{equation}

The resulting largest subspace would be part of $C^\un_c(\R[2n]\times\R[m],\Cc)$ and invariant with respect to $V$ as well as with respect to the Laplacians $\D_{\R[2n]}$ and $\D_{\R[m]}$ or any smooth partial differential operator in $C^\un_c(\R[m],\Cc)$.
\er
\section{The electro-weak interaction}\las{6}
The gauge group of the electro-weak interaction is $\SU(2)\times\U(1)$. To implement the $\U(1)$ action we have to use the $\SU(n+3)$ model with $n=1$. As noted in \rs{3} the $\SU(1+3)$ gauge field contains a general $\uc(1)$ connection.

For the realization of $\SU(2)$ we could either use the same method, i.e., looking at the $\SU(n+3)$ model with $n=2$, or use the $\suc(2)$ Lie subalgebra which is part of the $\SU(1+3)$ model as an embedding of $\suc(2)$ in $\suc(3)$, or we could simply use the fact that $SU(2)$ is the simply connected twofold cover of $\SO(3)$ and employ the corresponding gauge field which is known to be symmetric with respect to rigid motions of $\R[3]$.

The $SU(2+3)$ model has the disadvantage of the additional constraint equations, so this model should be avoided when possible. The remaining two possibilities are very similar. We shall choose the independent $\soc(3)$ realization of $\suc(2)$, which has already been used to define quantum cosmological models, \cf \cite{cg:qfriedman-ym,cg:qfriedman-ym2}. 

Let us briefly describe how $\soc(3)$ can be looked at as the Lie algebra of $\Ad(\SU(2))$.

Consider the standard generators $T_i$, $1\le i\le 3$, of $\soc(3)$ viewed as antisymmetric homomorphisms in $\R[3]$ such that
\begin{equation}
[T_i,T_j]=\e^k_{ij}T_k.
\end{equation}

Let $\mf g=\suc(2)$, then a basis of $i\mf g$ is given by the Pauli matrices $\s_i$, $1\le i\le 3$, satisfying
\begin{equation}
[\s_i,\s_j]=2i\e^k_{ij}\s_k.
\end{equation}

Now, the classical adjoint representation of $\SU(2)$ as homomorphisms of $\mf g$ gives just $\SO(3)$ and
\begin{equation}
\Ad_*(\tfrac1{2i}\s_k)=T_k,
\end{equation}
see e.g., \cite[Theorem 19.12]{frankel:book} and also \cite[equ. (1.12)]{faddeev:book}.

Note that $\Ad_*^{-1}$ is two-valued. Thus, let
\begin{equation}
\tilde A=\tilde\f T_a\om^a_i dx^i
\end{equation}
be an $\SO(3)$ connection, then it can be looked at as the adjoint connection of the $SU(2)$ connection
\begin{equation}
B=\tilde\f\tfrac1{2i}\s_k\om^k_i dx^i,
\end{equation}
where $\om^a$ is the form in \fre{3.5} for $\so=\R[3]$.

These connections can be extended to the spacetime by setting
\begin{equation}
\tilde A_0=B_0=0.
\end{equation}

The additional Lagrangian terms which have to be considered in the functional in \fre{4.1} are
\begin{equation}\lae{6.7}
\begin{aligned}
&\int_\Om\{\tfrac14 \tr(\tilde F_{\mu\lam}\tilde F^{\mu\lam})-\tfrac12 \bar\mu \ga_{ab}\bar g^{\mu\lam}A^a_\mu A^b_\lam\chi_0^{-\frac13}+\tfrac14 \tr(\hat F_{\mu\lam}\hat F^{\mu\lam})\\
& -\tfrac12[\tilde L_iE^\mu_a\ga^a D_\mu L^i +\tilde e_R E^\mu_a\ga^a D_\mu e_R+ \overline{\tilde L_iE^\mu_a\ga^a D_\mu L^i +\tilde e_R E^\mu_a\ga^a D_\mu e_R} ]\\
&-\tfrac12 \bar g^{\mu\lam}D_\mu\f \overline{D_\lam{\f}}\chi_0^{-\frac13}  -h_e(\bar\f_i\bar e_{R\al}L^{i\al}+\f_i\bar L^{i\al}e_{R\al})\chi_0^{-\frac16}-\hat U(\f)\chi_0^{-\frac23}\},
\end{aligned}
\end{equation}
where
\begin{equation}
\hat U(\f)=-m_1^2\abs{\f}^2+b_0\abs{\f}^4,\q b_0>0.
\end{equation}

$(\tilde F_{\mu\lam})$ is the field strength of the $\SU(2)$ adjoint connection $(\tilde A_\mu)$, which we write in the form
\begin{equation}
\tilde A_\mu=A_\mu+\bar A_\mu,
\end{equation}
where $\bar A_\mu$ is the flat connection, hence $A_\mu=(A^a_\mu)$ is a tensor; $\ga_{ab}$ is the Cartan-Killing tensor of the Lie algebra. The corresponding term in the functional represents the mass of the connection: $\bar\mu$ is called the mass of the connection $\tilde A_\mu$, \cf \cite[p. 2]{cg:qfriedman-ym}. 

$(\hat F_{\mu\lam})$ is the field strength of the $\SU(1+3)$ connection. We now  denote the connection by $C$ instead of $A$ and consequently $\hat C$ will be the effective $\U(1)$ connection.

With respect to the Dirac terms, the Higgs field and the Yukawa terms we roughly follow the definitions and notations in \cite[p. 201]{faddeev:book}, see also \cite{weinberg:salam}.

From \cite[equ. (3.15)]{cg:qfriedman-ym} we obtain
\begin{equation}
\begin{aligned}
&\tfrac14 \tr(\tilde F_{\mu\lam}\tilde F^{\mu\lam})-\tfrac12 \bar\mu \ga_{ab}\bar g^{\mu\lam}A^a_\mu A^b_\lam\bar g^{\mu\lam}\chi_0^{-\frac13}=\\
&\q 3\dot{\tilde\f}w^{-2}e^{-2f}-3\tilde\f^4e^{-4f}-3\bar\mu\tilde\f^2e^{-4f},
\end{aligned}
\end{equation}
where we have to set $\tilde\f=\f$, $\tilde \ka=0$ and $\bar\mu=-\mu$, when comparing the reference with the present situation.

The value of
\begin{equation}
\tfrac14 \tr(\hat F_{\mu\lam}\hat F^{\mu\lam})
\end{equation}
we infer from \re{3.20} and \fre{3.23}, noting that now $n=1$.

Before we inspect the Higgs field $\f=(\f^1,\f^2)$, let us look at the Dirac term.

Now, we use a different spinor basis such that
\begin{equation}
\ga^0=i\begin{pmatrix}
0&\I\\[\cma]
\I&0
\end{pmatrix},
\end{equation}
and the helicity operator $\ga^5$ is represented as 
\begin{equation}
\ga^5=-\ga^0\ga^1\ga^2\ga^3=i\begin{pmatrix}
\I&0\\[\cma]
0&-\I
\end{pmatrix},
\end{equation}
i.e., writing a spinor $\psi$ in the form
\begin{equation}
\psi=\begin{pmatrix}
\chi\\
\h
\end{pmatrix},
\end{equation}
then $\chi=(\chi_\al)$, $1\le\al\le 2$, is left-handed and $\h=(\h_\bet)$, $1\le\bet\le 2$, is right-handed.

The Dirac terms in \re{6.7} have to be understood as inserting
\begin{equation}\lae{6.15}
L^i\q\ra\q \begin{pmatrix}
L^i\\
0
\end{pmatrix},\q 1\le i\le 2,
\end{equation}
and
\begin{equation}\lae{6.16}
e_R\q\ra\q \begin{pmatrix}
0\\
e_R
\end{pmatrix},
\end{equation}
where $L^i$ and $e_R$ are Weyl spinors
\begin{equation}
L^i=(L^i_\al)\q\wed\q e_R=(e_{R\bet}).
\end{equation}

The covariant derivatives of $L^i$ \resp $e_R$ are defined by
\begin{equation}\lae{6.18}
D_\mu L^i=L^i_{,\mu}+\C_\mu L^i+g_2B_\mu L^i+\tfrac12 g_3\hat C_\mu L^i
\end{equation}
and
\begin{equation}
D_\mu e_R=e_{R,\mu}+\C_\mu e_r + g_3 \hat C_\mu e_R,
\end{equation}
where $g_2$, $g_3$ are positive coupling constants. Note, that, whenever $L^i$ or $e_R$ are acted upon by the Dirac matrices $\ga^a$, then they have to be expressed in the form \re{6.15} \resp \re{6.16}, while, when acted upon by the Pauli matrices, they are simply Weyl spinors.

The terms
\begin{equation}
\hat C_\mu L^i\q\wed\q \hat C_\mu e_R
\end{equation}
are defined by using the convention in \fre{3.13} as well as the remarks following \fre{3.27}, hence
\begin{equation}
\hat C_k=0,\qq 1\le k\le 3,
\end{equation}
and
\begin{equation}
\hat C_0 L^i=i\vartheta L^i,\qq\vt\in\R[].
\end{equation}

Let us write \re{6.18} explicitly in terms of
\begin{equation}
\begin{pmatrix}
L^i\\
0
\end{pmatrix}\q\wed\q L^i,
\end{equation}
\begin{equation}
\begin{aligned}
D_\mu \begin{pmatrix}
L^i\\
0
\end{pmatrix}=\begin{pmatrix}
L^i_{,\mu}\\
0
\end{pmatrix}+\C_\mu \begin{pmatrix}
L^i\\
0
\end{pmatrix}+ g_2\begin{pmatrix}
B_\mu L^i\\
0
\end{pmatrix}+\tfrac{g_3}2\begin{pmatrix}
i\vt L^i\\
0
\end{pmatrix},
\end{aligned}
\end{equation}
and similarly for $e_R$.

Applying the definitions of $\ga^0$, $\ga^k$ we then deduce, by replacing at the end of the computation
\begin{equation}
L^i\q\ra\q L^i e^{\frac32 f}
\end{equation}
and
\begin{equation}
e_R\q\ra\q e_R e^{\frac32 f}
\end{equation}
without changing the notation,
\begin{equation}\lae{6.27}
\tilde L^i E^\mu_a\ga^a D_\mu L_i=-i\bar L^\al_i\tfrac D{dt}L^i_\al w^{-1}e^{-3f}+ \tfrac32 g_2\tilde \f \bar L^\al_i L^i_\al e^{-4f}
\end{equation}
and
\begin{equation}
\tilde e_R E^\mu_a\ga^a D_\mu e_R=-i \bar e^R_\al \tfrac D{dt}e_{R\al} w^{-1}e^{-3f},
\end{equation}
where
\begin{equation}
\tfrac D{dt}L^i_\al =L^i_{\al,t}+\tfrac{g_3}2i\vt L^i_\al
\end{equation}
and
\begin{equation}
\tfrac D{dt} e_{R\al}=\dot e_{R\al}+g_3i\vt e_{R\al}.
\end{equation}

Let us now consider the Higgs field $\f=(\f^i(t))$, $1\le i\le 2$. Its covariant derivative is defined by
\begin{equation}
D_\mu\f=\f_{,\mu}+g_2B_\mu\f+\tfrac{g_3}2\hat C_\mu\f,
\end{equation}
hence
\begin{equation}
D_0\f=\dot\f+\tfrac{g_3}2i\vt\f,
\end{equation}
\begin{equation}
D_k\f=-i\tfrac {g_2}2 \tilde\f\s_k\f,
\end{equation}
and
\begin{equation}
-\tfrac12 \bar g^{\mu\lam} D_\mu\f \overline{D_\lam\f}=\tfrac12w^{-2}\tfrac D{dt}\f\overline{\tfrac D{dt}\f}-\tfrac32 g_2^2\tilde\f^2\abs\f^2e^{-2f}.
\end{equation}

Writing the complex functions $\f^i$ as
\begin{equation}
\f^i=a^i+ib^i,
\end{equation}
we infer
\begin{equation}
\begin{aligned}
\bar\f_i\bar e_{R\al}L^{i\al}+\f_i\bar L^{i\al}e_{R\al}&=-a_i(\bar e_{R\al}L^{i\al}+\bar L^{i\al}e_{R\al})\\
\hp=&\q \q\q-b_i(i\bar L^{i\al}e_{R\al}-i\bar e_{R\al}L^{i\al}),
\end{aligned}
\end{equation}
hence, after quantization, it will be a self-adjoint operator in the finite dimensional Hilbert space generated by the fermions. However, the operator will depend on the spatial variables $a_i$, $b_i$, which will turn out to have very important consequences.

Note that a similar term appears on the right-hand side of \re{6.27}, i.e., even without the Yukawa term there would be a self-adjoint operator in the antisymmetric Fock space depending on the spatial variables---for the consequences we refer to \frr{11.5}.

The constants $g_2$, $g_3$, $b_0$ and $h_e$ are assumed to be positive, while $m_1$ may be real or imaginary. Note that the sign of $h_e$ is irrelevant.

\section{Quantization of the full Lagrangian}\las{7}
Adding the terms in \re{6.7} to the functional $J$ in \fre{4.1} and following the procedures in \rs{4} we arrive at an analogue of equation \fre{4.32} which reads
\begin{equation}\lae{7.1}
\begin{aligned}
J&=\int_a^bw\{G_{AB}\tfrac D{dt}y^A\tfrac D{dt}y^Bw^{-2}-2\al_M^{-1}\Lam e^{3f}-Ve^{-f}-Ue^{-f}\\
&\hp= -(3\tilde\f^4+3\bar\mu\tilde\f^2 +\tfrac32g_2^2\tilde\f^2\abs\f^2+\hat V+\hat U+\tfrac32 g_2\tilde\f \bar L^\al_iL^i_\al)e^{-f}\}\\
&\hp=+\int_a^b\{\frac{i}2(\bar\chi_i^a\tfrac D{dt}\chi^i_a+\bar L^\al_i\tfrac D{dt}L^i_\al+\bar e_R^\al\tfrac D{dt}e_{R\al})+\tup{c.c.}\\
&\q -m(\bar\chi^{\underline a}_i\chi^i_{\underline a}-\bar\chi^{\bar a}_i\bar\chi^i_{\bar a})e^{-f}w-h_e\big(a_i(\bar e_{R\al}L^{i\al}+\bar L^{i\al}e_{R\al})\\
&\hp= \qq\qq\qq\qq\q+b_i(-i\bar e_{R\al}L^{i\al}+i\bar L^{i\al}e_{R\al})\big)e^{-f}w\},
\end{aligned}
\end{equation}
 where
 \begin{equation}
\hat V=\hat\f_0^4+8\hat\f_0^2\abs{\hat z}^2+\abs{\hat z}^4,
\end{equation}
$\hat z\in\Cc$, is the potential coming from the energy of the connection $C_\mu$, and where
\begin{equation}
G_{AB}\tfrac D{dt}y^A\tfrac D{dt}y^B
\end{equation}
has now been modified to incorporate the new variables. Note also that the covariant derivative \cq{$\tfrac D{dt}$} is defined differently depending on the variables it is applied to.

The variable $y=(y^A)$ is now defined by
\begin{equation}
(y^A)=(f,\underset{\SU(n)}{\underbrace{\f_0,z^i,\F^i}},\underset{\SU(2)\times \U(1)}{\underbrace{\tilde\f,\hat\f_0,\hat z,\f^i}}).
\end{equation}
The additional variables are the real variables $\tilde\f$, $\hat\f_0$, the complex variable $\hat z$, and 
\begin{equation}
\f=(\f^i)\in \Ccc[2].
\end{equation}

Let us summarize the definitions of the covariant derivatives for the additional variables
\begin{equation}\lae{7.6}
\tfrac D{dt}\hat z=\hat z_{,t}+\tfrac 43i\vt \hat z,
\end{equation}
\cf \fre{3.23},
\begin{equation}\lae{7.7}
\tfrac D{dt}\f=\dot\f+\tfrac{g_3}2 i\vt\f,
\end{equation}
\begin{equation}\lae{7.8}
\tfrac D{dt}L^i_\al=L^i_{\al,t}+\tfrac{g_3}2i\vt L^i_\al,
\end{equation}
and
\begin{equation}\lae{7.9}
\tfrac D{dt}e_{R\al}=\dot e_{R\al}+g_3i\vt e_{R\al}.
\end{equation}

The metric $(G_{AB})$ is the diagonal Lorentz metric 
\begin{equation}
(G_{AB})=\diag(-\al_M^{-1}12e^{2f},12,6\I_{2n},\I_{2n},6,12,6\I_{2},\I_4)e^f.
\end{equation}

Canonical quantization then leads to the Wheeler-DeWitt equation
\begin{equation}
H\Psi=0,
\end{equation}
where the Hamilton operator $H$ is defined by
\begin{equation}
\begin{aligned}
e^f H&=-\tfrac12e^f\D+2\al_M^{-1}\Lam e^{4f}+V+U+\hat V+\hat U\\
&\hp= \;+(3\tilde\f^4+3\bar\mu\tilde\f^2 +\tfrac32g_2^2\tilde\f^2\abs\f^2+\tfrac32 g_2\tilde\f \bar L^\al_iL^i_\al)\\
&\q +m(\bar\chi^{\underline a}_i\chi^i_{\underline a}-\bar\chi^{\bar a}_i\chi^i_{\bar a})+h_e\big(a_i(\bar e_{R\al}L^{i\al}+\bar L^{i\al}e_{R\al})\\
&\hp= \qq\qq\qq\q+b_i(-i\bar e_{R\al}L^{i\al}+i\bar L^{i\al}e_{R\al})\big),
\end{aligned}
\end{equation}
and the Laplace operator with respect to the metric $(G_{AB})$ can be expressed as
\begin{equation}\lae{7.13}
-e^f\D\Psi=\frac {\al_M}{12}e^{-(2n+5)f}\frac\pa{\pa y^0}\Big(e^{(2n+3)f}\frac{\pa\Psi}{\pa y^0}\Big)-2a^{\al\bet}\Psi_{\al\bet},
\end{equation}
where
\begin{equation}
(a^{\al\bet})=\diag(\tfrac1{24},\tfrac1{12}\I_{2n},\tfrac12\I_{2n},\tfrac1{12},\tfrac1{24},\tfrac1{12}\I_2,\tfrac12\I_4).
\end{equation}
 Replacing $e^fH$ by $H$ without changing the notation, we then have
 \begin{equation}
H=H_1-H_0,
\end{equation}
 where
 \begin{equation}\lae{7.16}
H_0\Psi=-\frac {\al_M}{24}e^{-(2n+5)f}\frac\pa{\pa y^0}\Big(e^{(2n+3)f}\frac{\pa\Psi}{\pa y^0}\Big)-2\al_M^{-1}\Lam e^{4f}\Psi
\end{equation}
 and
\begin{equation}\lae{7.17}
\begin{aligned}
&H_1\Psi=\\
&-a^{\al\bet}\Psi_{\al\bet}+(V+U+\hat V+\hat U)\Psi
+(3\tilde\f^4+3\bar\mu\tilde\f^2 +\tfrac32g_2^2\tilde\f^2\abs\f^2)\Psi\\
&+m(\bar\chi^{\underline a}_i\chi^i_{\underline a}-\bar\chi^{\bar a}_i\chi^i_{\bar a})\Psi+\tfrac32 g_2\tilde\f \bar L^\al_iL^i_\al\Psi\\
&+h_e\big(a_i(\bar e_{R\al}L^{i\al}+\bar L^{i\al}e_{R\al})
+b_i(-i\bar e_{R\al}L^{i\al}+i\bar L^{i\al}e_{R\al})\big)\Psi.
\end{aligned}
\end{equation}
Note that the symbols $f,\f_0,z^i,\F^i,\tilde\f,\hat\f_0,\hat z^i,\f^i$ now are variables of the Euclidean space 
\begin{equation}
\R[]\times \R[4n+9],
\end{equation}
 where $f$ corresponds to the first factor. The complex variables have been expressed by their real and imaginary parts respectively, e.g.,
 \begin{equation}
\f_k=a_k+ib_k.
\end{equation}
 
 The terms in the last two rows  of the right-hand side of \re{7.17} represent a symmetric operator in the finite dimensional Hilbert space generated by the fermions which also depends on the spatial variables $a_k,b_k$ and $\tilde\f$.
 
 Let us write this operator in the form
 \begin{equation}
B+C,
\end{equation}
 where $B$ acts on the fermions from the $\SU(n)$ model and $C$ on those from the $\SU(2)\times\U(1)$ model, and let us abbreviate the rest of the right-hand side by $A$ such that
 \begin{equation}
H_1=A+B+C.
\end{equation}
 In the next section we shall define the Hilbert space in which $H_1$ acts as a symmetric operator.
 \section{The vector space defined by the constraints of the electro-weak interaction}\las{8}
 
 The functional in \fre{7.1} contains $\vt$ as a non-dynamical variable, hence an additional constraint equation has to be satisfied. The equations \re{7.6}--\fre{7.9} reveal how $\vt$ enters into the Lagrangian.
 
 Writing $\hat z$ \resp $\f^i$ in the form
 \begin{equation}
\hat z=\hat x+i\hat y
\end{equation}
 \resp
 \begin{equation}
\f^i=\xi^i+i\h^i
\end{equation}
 for $1\le i\le 2$, we deduce from \fre{5.35} that the differential oper\-a\-tor---we now use the notations $\lam_0,\tilde \lam_0$ and $\hat\lam_0$---has the form
 \begin{equation}
\lam_0=\tfrac43(\hat x\pde{}{\hat y}-\hat y\pde{}{\hat x})+i\tfrac83,
\end{equation}
 and a variant of \re{5.35} is also valid for $\f^i$, namely,
 \begin{equation}
g_3\tilde\lam_0=g_3\tfrac12(\xi^i\pde{}{\h^i}-\h^i\pde{}{\xi^i})+ig_3,
\end{equation}
 where, however, we now have to sum over $i$. The different coefficients are due to the different definitions of the covariant derivative, \cf \fre{7.7} and also \frr{5.4}---but note that we used the standard definitions.
 
 Finally, when differentiating the Dirac terms with respect to $-i\pde{}{\vt}$ we obtain
 \begin{equation}
g_3\hat\lam_0\equiv ig_3\hat\Lam_0=ig_3\{\tfrac12\bar L^\al_iL^i_\al+\bar e^\al_Re_{R\al}\},
\end{equation}
 where the summation convention is in place for all indices.
 
 Hence the constraint equation is
\begin{equation}
(\lam_0+g_3\tilde\lam_0+g_3\hat\lam_0)\Psi=0.
\end{equation}
 
 To solve this equation we first determine the eigenspaces of $\hat\lam_0$, or equivalently, of $\hat\Lam_0$, which is a self-adjoint operator in the $2^6$ dimensional Hilbert space $\mc F_2$ spanned by the electro-weak fermions. It has $9$ eigenvalues
 \begin{equation}
0,\tfrac12, \dots, \tfrac72, 4
\end{equation}
 which are all multiples of $\tfrac12$. This claim can be proved by arguing as in the proof of \frl{5.2a}.
 
 Denote by $\rho_a$, $1\le a\le 9$, these eigenvalues and by
 \begin{equation}
F_{\rho_a}
\end{equation}
 the corresponding eigenspaces, then
 \begin{equation}\lae{8.9}
\mc F_2=\bigoplus_{a=1}^9F_{\rho_a}.
\end{equation}

 Let $F_{\rho_a}$ be arbitrary. We shall use the operator $\tilde \lam_0$ to define a matching function space.
 
 \bt\lat{8.1}
 For any $r\in\Z$ there exists a largest infinite dimensional vector space
 \begin{equation}
E\su C^\un_c(\R[4],\Cc)
\end{equation}
such that all $u\in E$ satisfy
\begin{equation}
\tilde \lam_0u=-i\tfrac{r}2u,
\end{equation}
and such that $E$ will be invariant with respect to the operators $\D_{\R[4]}$ and 
\begin{equation}
u\ra Vu,
\end{equation}
where the potential $V$ is of the form
\begin{equation}
V=V_0(\abs z^2).
\end{equation}
The claims in \frr{5.5} are also valid.
 \et
 \bp
 The proof is similar to the proof of \frt{5.3} \resp \frt{5.8}.  First, let $\rho\in C^\un_c(\R[])$, then the functions
 \begin{equation}
\f=\rho(\abs\zeta^2),
\end{equation}
where $\zeta^i=\xi^i+i\h^i$, $1\le i\le 2$, satisfy
\begin{equation}
\tilde\lam_0\f=0.
\end{equation}

Second, let
\begin{equation}
u_k=\xi^k-i\h^k\q\wed\q \tilde u_k=\xi^k+i\h^k,
\end{equation}
$1\le k\le 2$ fixed, then
\begin{equation}
\tilde \lam_0u_k=-i\tfrac12 u_k +iu_k\q\wed\q \tilde \lam_0\tilde u_k=i\tfrac12 \tilde u_k +i\tilde u_k
\end{equation}

For $r\in\N$ define
\begin{equation}
u=u_k^r\rho(\abs\zeta^2)\q\wed\q \tilde u=\tilde u_k^r\rho(\abs\zeta^2)
\end{equation}
where $\rho\in C^\un_c(\R[])$ is arbitrary, then
\begin{equation}
\tilde\lam_0u=-i\tfrac{r}2u+iu\q\wed\q \tilde\lam_0\tilde u=i\tfrac{r}2\tilde u+i\tilde u
\end{equation}
and these functions, $u$ \resp $\tilde u$, generate an infinite dimensional subspace.

The invariance properties of the subspace can be proved as in the case of \rt{5.3}, and
the arguments at the end of the proof of that theorem  yield the existence of a largest subspace with these properties.
 \ep
 
 Next we have to define a function space $E_0$ such that 
 \begin{equation}
\lam_0 v=0\qq\A\, v\in E_0.
\end{equation}

This can be achieved with the help of \frt{5.8}. Let $E_0\su C^\un_c(\R[2],\Cc)$ be such that
\begin{equation}
(\hat x\pde{}{\hat y}-\hat y\pde{}{\hat x})v=-2iv\qq \A\, v\in E_0,
\end{equation}
then
\begin{equation}
\lam_0v=0\qq\A\,v\in E_0.
\end{equation}

 \section{The eigenvalue problem for the strong interaction}\las{9}
 In this section we want to solve the free eigenvalue problem for the matter Hamiltonian $H_{M_1}$ in the $\SU(n)$, $n\ge 2$, model. The Hamiltonian can be expressed in the form
\begin{equation}
\begin{aligned}
H_{M_1}\Psi&=(-a^{\al\bet}\Psi_{\al\bet}+V\Psi)+(-\tfrac12\D\Psi+U\Psi)+H_{F_1}\Psi\\
&\equiv H_1\Psi+H_2\Psi+H_{F_1}\Psi.
\end{aligned}
\end{equation} 
 The operator $H_1$ depends on the variables $(\f_0,z^i)\in\R[1+2n]$, $H_2$ on the variables $(\F^i)\in\R[2n]$ and $H_{F_1}$ acts on the fermions in a $2^{4n}$ dimensional subspace of the antisymmetric Fock space.
 
 Symbolizing the differentiation with respect to $\f_0$ by a prime and the Laplace operator with respect to $z\in \R[2n]$ by $\tilde\D$, then
 \begin{equation}
H_1\Psi=-\tfrac1{24}\Psi''-\tfrac1{12}\tilde\D\Psi+V(\f_0,z)\Psi.
\end{equation}
 \bd
 (i) To solve the eigenvalue problem for the operator $H_1$, we choose a largest subspace $E_1\su C^\un_c(\R[1+2n])$ the elements of which satisfy the constraint equations for the constrained operators $l_{k,m}$ and $l_k$ with eigenvalue $r=0$ and the invariance conditions, and define the Hilbert spaces
 \begin{equation}
\mc H_1=\bar E_1^{\norm\cdot}, 
\end{equation}
as the completion of $E_1$ in the $L^2$-norm, abbreviated simply by $\norm\cdot$, and $\tilde{\mc H}_1$ as the completion of $E_1$ with respect to the norm
\begin{equation}
\spd uu_1=\norm u_1^2=\int_{\R[]\times\R[2n]}(\abs {Du}^2+\abs x^4\abs u^2)
\end{equation}
where $x=(x^i)\in\R[1+2n]$.

\cvm
(ii) In case of the operator $H_2$, we first have to choose one of the joint eigenspaces $F_{\s_k}$ of the fermionic  constraint operators, \cf \frr{5.2}. Let $E_2=E_2(\s_k)$ be the matching largest subspace of $C^\un_c(\R[2n],\Cc)$ such that the constraint equations will be satisfied for
\begin{equation}
u\otimes\h,\qq \A\, (u,\h)\in (E_2\times F_{\s_k}).
\end{equation}
Then we define the Hilbert spaces $\mc H_2=\mc H_2(\s_k)$ as the completion of $E_2$ with respect to the $L^2$-norm
\begin{equation}
\spd uu=\norm u^2=\int_{\R[4n]}\abs u^2
\end{equation}
and $\tilde{\mc H}_2$ as the completion of $E_2$ with respect to the norm
\begin{equation}
\spd uu_1=\norm u_1^2=\int_{\R[2n]}(\abs{Du}^2+\abs{x}^{2p}\abs u^2),
\end{equation}
where $x=(x^i)\in\R[2n]$ and $p$ the exponent in \fre{3.34}. 
 \ed

 We then have to solve three eigenvalue problems for the Hamiltonians $H_i$ in $\mc H_i$, $1\le i\le 2$, and for the fermionic Hamiltonian $H_{F_1}$ restricted to $F_{\s_k}$. $H_{F_1}$ corresponds to a quadratic form, i.e., there holds 
 \begin{equation}
a(\xi,\h)=\spd{H_{F_1}\xi}\h \qq\A\,\xi,\h\in \mc F_1,
\end{equation}
 where $a$ is a hermitean bilinear form. In general the spaces $F_{\s_k}$ will not be invariant with respect to $H_{F_1}$---note, however, that the $16$ mutually orthogonal unit vectors given in \frl{5.7b} are all eigenvectors of $H_{F_1}$. We therefore define a new fermionic Hamiltonian operator $H_f=H_f(\s_k)$ as the unique self-adjoint operator $H_f\in L(F_{\s_k},F_{\s_k})$ satisfying
 \begin{equation}
a(\xi,\h)=\spd{H_f\xi}\h\qq\A\, \xi,\h\in F_{\s_k}.
\end{equation}
 Its eigenvectors will then complement the eigenvectors of the bosonic Hamiltonians.
 
 When solving the bosonic problems it suffices to look at just one operator, and we choose $H_2$ because the corresponding potential $U$ is more general and the proof slightly more elaborate.
 \bt
 The linear operator $H_2$ with
 \begin{equation}
D(H_2)=E_2\su\mc H_2
\end{equation}
is symmetric and semi-bounded from below. Let $\hat H_2$ be its self-adjoint Friedrichs extension, then there exist countably many eigenvectors 
\begin{equation}
u_i\in \tilde{\mc H}_2\hra \mc H_2
\end{equation} with eigenvalues $\lam_i$ of finite multiplicities of $\hat H$,
\begin{equation}\lae{9.12}
\hat H_2u_i=\lam_i u_i,
\end{equation}
satisfying
\begin{equation}\lae{9.13}
\spd{u_i}{u_j}=0\qq\A\, i\ne j,
\end{equation}
\begin{equation}\lae{9.14}
\lam_i \le \lam_{i+1}\q\wed\q \lim_{i\ra\un}\lam_i=\un.
\end{equation}
The $(u_i)$ are complete in $\tilde{\mc H}_2$ as well as in $\mc H_2$.
 \et
 \bp
 (i) We shall derive the existence of eigenfunctions from a general variational problem. The symmetric operator $H_2$ defines a sesquilinear form $a$
 \begin{equation}
a(u,v)=\spd{H_2 u}v=\int_{\R[2n]}\{\tfrac12 D_iuD^i\bar v+Uu\bar v\}\qq\A\, u.v\in D(H_2),
\end{equation}
where we used that
\begin{equation}
H_2u=-\tfrac12\D u+Uu\qq\A\,u\in D(H_2),
\end{equation}
and integrated by parts. In view of the estimates \fre{3.34} the quadratic form
\begin{equation}
a(u,u)+c_2\norm u^2
\end{equation}
is equivalent to
\begin{equation}
\spd uu_1.
\end{equation}

Furthermore, the norm $\norm\cdot$ is compact relative to $\norm\cdot_1$, i.e., if
\begin{equation}\lae{9.19}
u_i \rha u \qq\tup{in}\; \tilde{\mc H}_2,
\end{equation}
then
\begin{equation}\lae{9.20}
u_i\ra u \qq\tup{in}\; \mc H_2,
\end{equation}
where we used the trivial embedding
\begin{equation}
\tilde{\mc H}_2\hra \mc H_2;
\end{equation}
the property described in \re{9.19}, \re{9.20} can be rephrased that this embedding is compact.

The compactness proof is similar to the proof of \cite[Lemma 6.8]{cg:qfriedman}, where a one dimensional analogue has been considered, but the arguments in the higher dimensional case are the same.

A general variational argument which goes back to Courant-Hilbert, see e.g., \cite{cg:eigenwert}, then yields the existence of a mutually orthogonal sequence $(u_i)$ of eigenvectors solving the variational relation
\begin{equation}\lae{9.22}
a(u_i,v)=\lam_i\spd{u_i}v\qq\A\, v\in \tilde{\mc H}_2,
\end{equation}
such that the relations \re{9.13}, \re{9.14}  and the completeness claims in $\tilde{\mc H}_2$ as well as $\mc H_2$ are valid.

\cvm
(ii)  To prove \re{9.12} we consider the closure $\tilde H_2$ of $H_2$. Let $u\in D(\tilde H_2)$, then there exists a sequence $u_k\in D(H_2)$ such that
\begin{equation}
u_k\ra u\qq \tup{in}\; \mc H_2,
\end{equation}
and
\begin{equation}
H_2u_k\ra \tilde H_2u\qq\tup{in}\;\mc H_2.
\end{equation}

Define $f_k$ formerly by
\begin{equation}
f_k=H_2u_k.
\end{equation}
Multiplying the equation
\begin{equation}
H_2(u_k-u_l)=f_k-f_l
\end{equation}
by $(\bar u_k-\bar u_l)$ and integrating by parts we conclude
\begin{equation}
a(u_k-u_l,u_k-u_l)\le \norm{f_k-f_l}\msp[1]\norm{u_k-u_l},
\end{equation}
hence, $(u_k)$ is also a Cauchy sequence in $\tilde{\mc H}_2$, and we conclude further
\begin{equation}
D(\tilde H)\su \tilde{\mc H}_2.
\end{equation}

The Friedrichs extension $\hat H_2$ of $\tilde H_2$ is then defined by
\begin{equation}
\hat H_2=\fv{H_2^*}{D(H_2^*)\ii\tilde{\mc H}_2},
\end{equation}
where $H_2^*$ is the adjoint of $H_2$.
 
 Now, let $u_i$ be an arbitrary solution of \re{9.22}, then we deduce immediately
 \begin{equation}
u_i\in D(H_2^*)\q\wed\q H_2^*u_i=\lam_iu_i,
\end{equation}
 proving \re{9.12}.
 \ep
 
 A finite number of the eigenvalues $\lam_i$ of the variational solutions  can be negative, since the potential $U$ is not supposed to be non-negative, but only subject to the estimates in \fre{3.34}.
 
 The positivity of the smallest eigenvalue $\lam_0$ can be guaranteed under the following assumptions:
 
 \bt\lat{9.3}
 Let $c_1,c_2$ be the constants  in \re{3.34} and let $c_1$ be fixed, then there exists a positive constant $c_0$ such that the smallest eigenvalue $\lam_0$ of the variational problems \re{9.22} is strictly positive provided 
 \begin{equation}\lae{9.31}
c_2<c_0.
\end{equation}
Moreover, for fixed $c_2$, let
\begin{equation}
\lam_0=\lam_0(c_1)
\end{equation}
be the smallest eigenvalue, then
\begin{equation}\lae{9.33}
\liminf_{c_1\ra\un}\lam_0(c_1)=\un.
\end{equation}
 \et
 
 \bp
 (i) Let us first prove the positivity of $\lam_0$, if \re{9.31} is satisfied. The eigenfunction of the smallest eigenvalue $\lam_0$ is a solution of the variational problem
 \begin{equation}
J(v)=\int_{\R[2n]}(\tfrac12\abs{Dv}^2+U\abs v^2)\ra\min\qq\A\,v\in K,
\end{equation}
where
\begin{equation}
K=\set{v\in\tilde{\mc H}_2}{\norm v=1}.
\end{equation}

In view of \fre{3.34} $J$ can be estimated from below by
\begin{equation}\lae{9.36}
\int_{\R[2n]}(\tfrac12\abs{Dv}^2+c_1\abs x^{2p}\abs v^2-c_2\abs v^2).
\end{equation}

Denote by $\tilde J$ the functional
\begin{equation}
\tilde J(v)=\int_{\R[2n]}(\tfrac12\abs{Dv}^2+c_1\abs x^{2p}\abs v^2),
\end{equation}
then the variational problem
\begin{equation}
\tilde J(v)\ra \min\qq\A\,v\in K
\end{equation}
has a solution $\tilde u_0$ with eigenvalue $\tilde\lam_0>0$, i.e., there holds
\begin{equation}
0<\tilde\lam_0=\tilde J(\tilde u_0)\le \tilde J(v)\qq\A\,v\in K.
\end{equation}

Thus, setting
\begin{equation}
c_0=\tilde\lam_0
\end{equation}
will prove the first claim.

\cvm
(ii) To prove \re{9.33}, we argue by contradiction. Let $c_{1,k}$ be sequence converging to infinity and $u_k$ a corresponding sequence of first eigenfunctions such that
\begin{equation}
\lam_{0,k}\le\const\qq\A\,k.
\end{equation}

Hence, we have
\begin{equation}
J(u_k)=\lam_{0,k}=\lam_{0,k}\norm{u_k}^2.
\end{equation}
Since $c_2$ is fixed, we deduce from \re{9.36}
\begin{equation}
\int_{\R[2n]}(\tfrac12\abs{Du_k}^2+c_{1,k}\abs x^{2p}\abs{u_k}^2)\le \lam_{0,k}+c_2\le c.
\end{equation}

The sequence $(u_k)$ is therefore bounded in $\tilde{\mc H}_2$ and
\begin{equation}
\lim_{k\ra\un}\int_{\R[2n]}\abs x^{2p}\abs {u_k}^2=0,
\end{equation}
and we conclude, since the embedding
\begin{equation}
\tilde{\mc H}_2\hra \mc H_2
\end{equation}
is compact, that a subsequence, not relabeled, converges weakly in $\tilde{\mc H}_2$ to a function $u$ such that
\begin{equation}
u_k\ra u\qq\tup{in}\;\mc H_2;
\end{equation}
hence, $\norm u=1$ contradicting
\begin{equation}
\int_{\R[2n]}\abs x^{2p}\abs u^2\le \lim\int_{\R[2n]}\abs x^{2p}\abs {u_k}^2=0.
\end{equation}
 \ep
 
For the Hamiltonian $H_1$ similar results are valid. The potential $V$ then satisfies
\begin{equation}
c_1\abs x^4\le V,\qq c_1>0,
\end{equation}
 if $x=(x^i)\in \R[1+2n]$. Hence, the smallest eigenvalue $\lam_0$ is always positive, but we cannot manipulate its size, since we cannot adjust $V$.
 
 Combining the results for the Hamiltonians $H_1$, $H_2$, and $H_{F_1}$ we have proved:
\bt\lat{9.4}
For each $F_{\s_k}\su\mc F_1$, $1\le k\le 9$, there exist infinite dimensional Hilbert spaces $\mc H_1$ and $\mc H_2$ and corresponding self-adjoint operators $\hat H_1, \hat H_2$ and $H_f$ in $F_{\s_k}$, such that the functions in
\begin{equation}
\mc H_1\otimes \mc H_2\otimes F_{\s_k}
\end{equation}
satisfy the constraint equations, and complete sequences of eigenfunctions
\begin{equation}
u_i\in\mc H_1\q\wed\q v_j\in\mc H_2
\end{equation}
for $\hat H_1$ \resp $\hat H_2$ and finitely many eigenvectors for $H_f$
\begin{equation}
\h_l\in F_{\s_k}.
\end{equation}
The products
\begin{equation}
\Psi_{ijl}=u_i\otimes v_j\otimes\h_l
\end{equation}
are then eigenfunctions of 
\begin{equation}
\hat H_1+\hat H_2+H_f.
\end{equation}
Relabeling the eigenvalues and eigenfunctions we get a sequence of eigenvalues $\lam_i$ and corresponding eigenfunctions $\Psi_i$ such that
\begin{equation}
0<\lam_i\le \lam_{i+1}\q\wed\q \lim\lam_i=\un,
\end{equation}
\begin{equation}
\hat H_2\Psi_i=\lam_i\Psi_i,
\end{equation}
where, by abusing the notation, we define
\begin{equation}
\hat H_2=\hat H_1+\hat H_2+H_f,
\end{equation}
and
\begin{equation}
D(\hat H_2)=\langle(\Psi_i)_{i\in\N}\rangle.
\end{equation}
$\hat H_2$ is then essentially self-adjoint in 
\begin{equation}
\mc H_2=\mc H_1\otimes \mc H_2\otimes F_{\s_k}.
\end{equation}

 \et
 
 \section{The eigenvalue problem for the electro-weak interaction}\las{10}
 The matter Hamiltonian of the electro-weak interaction is the sum of two Hamiltonians which are strongly coupled and cannot be treated separately.
 \begin{equation}
H_{M_2}=H_3+H_{\mc F_2}
\end{equation}
The bosonic variables are $(\tilde\f,\hat\f_0,\hat z,\f^i)$, where $\tilde\f,\hat\f_0$ are real variables, $\hat z$ complex and $(\f^i)$ a complex doublet, the Higgs field. Only $\hat z$ and $\f^i$ are related with the constraint equations.

Let us denote the coordinates according to
\begin{equation}\lae{10.2}
(\tilde\f,\hat\f_0,\hat z,\f^i)\ra (x,y,\hat x+i\hat y,\xi^i+i\h^i).
\end{equation}
and the Laplacians in $\R[2]$ \resp $\R[4]$ by $\tilde \D$ \resp $\bar\D$.

With these notations there holds
\begin{equation}\lae{10.3}
\begin{aligned}
H_3\Psi&=-\tfrac1{12}\frac{\pa^2\Psi}{\pa x^2}-\tfrac1{24}\frac{\pa^2\Psi}{\pa y^2}-\tfrac1{12}\tilde\D\Psi-\tfrac12\bar\D\Psi+\hat V+\hat U\\
&\hp= \;+3x^4+3\bar\mu x^2 +\tfrac32g_2^2x^2(\abs\xi^2+\abs\h^2)+\tfrac32 g_2x\bar L^\al_iL^i_\al\\
&\q +h_e\big(\xi_i(\bar e_{R\al}L^{i\al}+\bar L^{i\al}e_{R\al})\\
&\hp= \qq\qq\qq\q+\h_i(-i\bar e_{R\al}L^{i\al}+i\bar L^{i\al}e_{R\al})\big),
\end{aligned}
\end{equation}
where $1\le\al\le 2$, $1\le i\le 2$.

The potential $\hat V$ is defined by
\begin{equation}\lae{10.4}
\hat V=\abs y^4+8y^2(\hat x^2+\hat y^2)+(\hat x^2+\hat y^2)^2,
\end{equation}
and $\hat U$ by
\begin{equation}\lae{10.5}
\hat U=b_0(\abs\xi^2+\abs\h^2)^2-m_1^2(\abs\xi^2+\abs\h^2),
\end{equation}
where $b_0>0$ and $m_1$ can be real or imaginary.

Let $\bar V$ be the potential
\begin{equation}\lae{10.6}
\bar V=3x^4+3\bar\mu x^2 +\tfrac32g_2^2x^2(\abs\xi^2+\abs\h^2),
\end{equation}
then we see that the sum of all three potentials has the same structure as the potentials in the case of  the strong interaction, namely,
\begin{equation}
-c_2+c_1\abs x^4\le \bar V+\hat V+\hat U\le c_1'\abs x^4+c_2',
\end{equation}
where  $x\in\R[8]$---but this usage is restricted to this particular estimate.

We also see that the fermionic operators have coefficients depending on $(x,\xi^k,\h^k)$ and therefore the eigenvalue problem cannot be separated in bosonic and fermionic parts but has to treated in a fermions valued function space. The eigenfunctions will be non-trivial fermionic fields
\begin{equation}
\Psi:\R[8]\ra \mc F_2,
\end{equation}
where $\mc F_2$ is the subspace of the antisymmetric Fock space spanned by the fermions.

$H_3$ is obviously formerly self-adjoint and the eigenvalues of the fermionic operators---disregarding their coefficients as well as $g_2$ and $h_e$---are absolutely bounded by a numerical constant $\al_0$.

Thus, using the symbol $u$ instead of $\Psi$, if
\begin{equation}
u,v\in C^\un_c(\R[8],\mc F_2)
\end{equation}
then
\begin{equation}
\spd{H_3u}v=\spd u{H_3v}
\end{equation}
and
\begin{equation}\lae{10.11}
\begin{aligned}
\spd{H_3u}u&=\int_{\R[8]}\{a^{ij}\spd{D_iu}{D_ju}+(\bar V+\hat V+\hat U)\norm{u}^2\\
&\hp= +\tfrac32 g_2xa_0(u,u)+h_e(\xi^ka_k(u,u)+\h^kb_k(u,u))\}, 
\end{aligned}
\end{equation}
where
\begin{equation}
-a^{ij}D_iD_j u
\end{equation}
represents the elliptic main differential part of $H_3$, and $a_0$, $a_k$, $b_k$, $1\le k\le 2$, are the sesquilinear fermionic forms, e.g.,
\begin{equation}
a_0=\tfrac12 \bar L^\al_kL^k_\al,
\end{equation}
and the scalar product under the integral sign is the scalar product in $\mc F_2$ with corresponding norm $\norm\cdot$.

Let $\chi\in\mc F_2$ be normalized, $\norm\chi=1$, then
\begin{equation}
\max(\abs{a_0(\chi,\chi)},\abs{a_k(\chi,\chi)},\abs{b_k(\chi,\chi)})\le \al_0\q\A\, 1\le k\le 2,
\end{equation}
and we deduce, that for any $\de>0$
\begin{equation}\lae{10.15}
\begin{aligned}
\spd{H_3u}u&\ge \int_{\R[8]}\{a^{ij}\spd{D_iu}{D_ju}+(\bar V+\hat V+\hat U)\norm{u}^2\\
&\hp\ge -c\big(g_2^2\abs x^2+h_e^2(\abs\xi^2+\abs\h^2)\big)\al_0^2\de^{-1}\norm u^2-\de\norm u^2\},
\end{aligned}
\end{equation}
where $c$ is a numerical constant.

Note that $u$ has values in $\mc F_2$, i.e., if we fix an orthonormal basis in $\mc F_2$,
\begin{equation}
u=(u^A),
\end{equation}
then
\begin{equation}
a^{ij}\spd{D_iu}{D_ju}=a^{ij}D_iu^AD_j\bar u_A,
\end{equation}
and
\begin{equation}
c_1\norm{Du}^2\le a^{ij}\spd{D_iu}{D_ju}\le c_2\norm{Du}^2,
\end{equation}
where $c_1$, $c_2$ are positive numerical constants, and the norm is the norm in $\mc F_2$.

To solve the eigenvalue problem we first have to define the Hilbert space. Fix an eigenspace $F_{\rho_a}$, $1\le a\le 9$, of $\hat\lam_0$ in $\mc F_2$, \cf \fre{8.9}, and let $E_0\su C^\un_c(\R[4],\Cc)$ \resp $E\su C^\un_c(\R[4],\Cc)$ be matching subspaces, \cf \frt{8.1} and the remarks at the end of \rs{8}. Then we define 
\begin{equation}
E=E(\rho_a)=E_0\otimes E\otimes F_{\rho_a}
\end{equation}
and consider $E$ as a subspace of $C^\un_c(\R[8],F_{\rho_a})$
\begin{equation}
E\su C^\un_c(\R[8],F_{\rho_a}),
\end{equation}
where its elements are functions
\begin{equation}
u=u(x)=(u^A)
\end{equation}
with pointwise norm
\begin{equation}
\norm u^2=u^A\bar u_A.
\end{equation}
\bd
Let $\mc H_3$ be the completion of $E$ with respect to the $L^2$-norm, where we define for $u\in E$
\begin{equation}
\norm u^2=\int_{\R[8]}\norm u^2;
\end{equation}
the norm inside the integral is the norm in $\mc F_2$. 

The Hilbert space $\tilde{\mc H}_3$ is defined as the completion of $E$ with respect to the norm
\begin{equation}
\norm u^2_1=\int_{\R[8]}\{\norm{Du}^2+\abs x^4\norm u^2\}.
\end{equation}
\ed

Though $E$ is invariant with respect to the potentials and the respective Laplace operators it is not invariant with respect to $H_3$ because of the fermionic operators which also depend on spatial variables. To define a meaningful symmetric operator satisfying the constraints, we consider the quadratic form associated with $H_3$ which is defined in \re{10.11}. Denote this quadratic form by $a_3$,
\begin{equation}
a_3(u,v)=\spd{H_3u}v\qq\A\, u,v\in E.
\end{equation}
In view of the estimate in \re{10.15}, $a_3$ is semi-bounded from below in $\mc H_3$, or more precisely, we have
\begin{equation}
a_3(u,u)\ge c_1\norm u^2_1-c_2\norm u^2\qq\A\,u\in E;
\end{equation}
 for a proof simply choose the parameter $\de$ in \re{10.15} large enough.
 
 On the other hand, $a_3$ can be estimated from above by
 \begin{equation}
a_3(u,u)\le c_1'\norm u^2_1+c'_2\norm u^2\le c_1'\norm u^2_1\qq\A\,u\in E,
\end{equation}
where the second inequality is valid because of the embedding
\begin{equation}
\tilde{\mc H}_3\hra \mc H_3
\end{equation}
is compact; the constant $c_1'$ in the second inequality is of course different from the corresponding constant in the first inequality.

Thus, $a_3$ has a natural extension to $\tilde{\mc H}_3$ and we can apply the general variational principle to find a complete set of eigenfunctions.
\bt\lat{10.2}
There exists a sequence of normalized eigenfunctions $u_i$ with real eigenvalues $\lam_i$ of finite multiplicities such that
\begin{equation}
a_3(u_i,v)=\lam_i\spd{u_i}v\qq\A\,v\in\tilde{\mc H}_3,
\end{equation}
\begin{equation}
\lam_i\le\lam_{i+1}\q\wed\q \lim\lam_i=\un,
\end{equation}
and
\begin{equation}
a_3(u_i,u_j)=\spd{u_i}{u_j}=0\qq\A\,i\ne j.
\end{equation}
Define the linear operator $T_3$ by
\begin{equation}
D(T_3)=\langle  (u_i)_{i\in\N}\rangle \q\wed\q T_3u_i=\lam_iu_i\q\A\,i\in\N,
\end{equation}
then $T_3$ is densely defined in $\mc H_3$, symmetric, essentially self-adjoint and there holds
\begin{equation}\lae{10.33}
a_3(u,v)=\spd{T_3u}v\qq\A\,u,v\in D(T_3).
\end{equation}
\et
\bp
We only have to prove the claims about the operator $T_3$. $T_3$ is certainly densely defined and satisfies \re{10.33}, since this relation is valid for $u=u_i$

Hence $T_3$ is symmetric and it remains to prove the essential self-adjointness. Thus it suffices to prove
\begin{equation}
\overline{R(T_3\pm i)}=\mc H_3.
\end{equation}
But these relations are obviously valid, since
\begin{equation}
u_i\in R(T_3\pm i)\qq\A\, i.
\end{equation}
\ep

The closure of $T_3$ is then the self-adjoint operator we are looking for
\begin{equation}
\hat H_3=\hat H_3(\rho_a)=\bar T_3.
\end{equation}

As in the case of the strong interaction a finite number of eigenvalues could be negative. This can be excluded by adjusting the free parameters $\bar\mu$ and $b_0$ in the potentials $\bar V$ and $\hat U$ appropriately.

Using the notations in \re{10.2}, \re{10.3} and the definitions of the potentials $\bar V$, $\hat V$, $\hat U$ in \re{10.6}, \re{10.4}, \re{10.5} we infer 
\begin{equation}
\begin{aligned}
\bar V+\hat V+\hat U&\ge 3\bar\mu \abs x^2 +b_0(\abs\xi^2+\abs\h^2)^2\\
&\hp\ge +3\abs x^4+\abs y^4+(\hat x^2+\hat y^2)^2-m_1^2(\abs\xi^2+\abs\h^2),
\end{aligned}
\end{equation}
and we conclude further, in view of \re{10.15},
\begin{equation}
\begin{aligned}
&a_3(u,u)\ge \\
&\int_{\R[8]}\{c_1\norm{Du}^2+\big((b_0-\tfrac{c^2}2h_e^4\al_0^4\de^{-3}-\tfrac{\abs m_1^4}2\de^{-1})(\abs\xi^2+\abs\h^2)^2\\
&\qq\qq+3\abs x^4+\abs y^4+(\hat x^2+\hat y^2)^2-2\de\big)\norm u^2\},
\end{aligned}
\end{equation}
provided
\begin{equation}\lae{10.39}
3\bar\mu\ge cg_2^2\al_0^2\de^{-1}.
\end{equation}

Hence, we conclude, as in the proof  of  \frt{9.3}: 
\bt\lat{10.3}
There exists a constant $\de=\de(c_1)>0$ such that the ei\-gen\-val\-ues $\lam_i$ are strictly positive provided 
\begin{equation}
b_0\ge \tfrac{c^2}2h_e^4\al_0^4\de^{-3}+\tfrac{\abs m_1^4}2\de^{-1}+1
\end{equation}
and $\bar\mu$ satisfies \re{10.39}.
\et

We have thus solved the eigenvalue problem for each subspace $F_{\rho_a}\su\mc F_2$ in a corresponding Hilbert space
\begin{equation}
\mc H_3(\rho_a).
\end{equation}
These Hilbert spaces are mutually orthogonal subspaces of
\begin{equation}
L^2(\R[8])\otimes \mc F_2\cong L^2(\R[8],\mc F_2).
\end{equation}
The self-adjoint operators $\hat H_3(\rho_a)$ then define a unique self-adjoint operator $\hat H_3$ in
\begin{equation}
\bigoplus_{a=1}^9\mc H_3(\rho_a)
\end{equation}
such that
\begin{equation}
\fv{\hat H_3}{\mc H_3(\rho_a)}=\hat H_3(\rho_a)\qq\A\, 1\le a\le 9.
\end{equation}

\section{The spectral resolution}\las{11}
We shall now prove the spectral resolution of the Wheeler-DeWitt equation for the full Hamiltonian when gravity is combined with the strong and electro-weak interactions. Our proof will even be valid when a finite number of matter fields are involved. However, except for the actual proof, we shall only consider the two interactions we are dealing with to simplify the presentation.

For arbitrary but fixed $\s_k$, $\rho_a$, $1\le a,k\le 9$, let $\mc H_2(\s_k)$, $\mc H_3(\rho_a)$ be the corresponding Hilbert spaces and $\hat H_2$ \resp $\hat H_3$ the (essentially) self-adjoint operators solving the eigenvalue problems
\begin{equation}
\hat H_2u_i=\lam_i u_i\qq u_i\in \mc H_2,
\end{equation}
\resp
\begin{equation}
\hat H_3 v_j=\mu_jv_j\qq v_j\in \mc H_3,
\end{equation}
\cf \frt{9.4} \resp \frt{10.2}.

The functions
\begin{equation}
\f_{ij}=u_i\otimes v_j\in \mc H_2\otimes\mc H_3
\end{equation}
are then eigenfunctions of the operator
\begin{equation}
\hat H_1=\hat H_2+\hat H_3,
\end{equation}
\begin{equation}
\hat H_1 \f_{ij}=(\lam_i+\mu_j)\f_{ij},
\end{equation}
where
\begin{equation}
D(\hat H_1)=\langle(\f_{ij})_{(i,j)\in\N\times\N}\rangle.
\end{equation}

We require that
\begin{equation}
\lam_i+\mu_j>0\qq\A\, (i,j).
\end{equation}

In view of the results in \frt{9.3} and \frt{10.3} this can always be achieved by choosing the parameters in the potentials appropriately.

After relabeling the countably many eigenvalues and eigenfunctions we may assume that $(\f_i,\mu_i)$ are solutions of the eigenvalue problem for $\hat H_1$ satisfying
\begin{equation}
\hat H_1\f_i=\mu_i\f_i
\end{equation}
such that the $(\f_i)$ are complete in $\mc H_1=\mc H_2 \otimes \mc H_3$ and the eigenvalues $\mu_i$ have finite multiplicities such that
\begin{equation}
0<\mu_i\le \mu_{i+1}\q\wed\q \lim\mu_i=\un.
\end{equation}

We also note that the elements $\f\in \mc H_1$ are viewed as maps
\begin{equation}
\f:\R[4n+9]\ra F_{\s_k}\otimes F_{\rho_a}\su \mc F_1\otimes\mc F_2,
\end{equation}
i.e.,
\begin{equation}
\mc H_1\su L^2(\R[4n+9],\mc F_1\otimes\mc F_2)
\end{equation}
We are therefore in a similar situation as in \cite{cg:qfriedman-ym2}, where we considered a related problem.

The Wheeler-DeWitt equation can now be written in the form
\begin{equation}
H_0\Psi-\hat H_1\Psi=0,
\end{equation}
where $\Psi$ has to satisfy the constraints. The constraints will be satisfied, if we split $\Psi$ in the form
\begin{equation}
\Psi=u\otimes\f,
\end{equation}
where $\f\in\mc H_1$ and $u$ is a complex valued function
\begin{equation}
u=u(f)\equiv u(t)
\end{equation}
depending on the real variable $f$ which we shall also denote by $t$.

The operator $H_0$ is the differential operator
\begin{equation}
H_0u=-\frac {\al_M}{24}e^{-(2n+5)t}\Big(e^{(2n+3)t}u'\Big)'-2\al_M^{-1}\Lam e^{4t}u
\end{equation}
\cf \fre{7.16}, where a dot or prime indicates differentiation with respect to $t$.

The exponents $(2n+3)$ \resp $(2n+5)$ depend on the number of the bosonic dynamical variables. To solve the Wheeler-deWitt equation for an arbitrary number of matter fields with $m$ dynamical bosonic variables, we consider the operator
\begin{equation}
H_0u=-\frac {\al_M}{24}e^{-\tfrac{(m+1)}2t}\Big(e^{\tfrac{(m-3)}2t}u'\Big)'-2\al_M^{-1}\Lam e^{4t}u.
\end{equation}
In our present situation there holds
\begin{equation}
m=4n+9.
\end{equation}

Let $\tilde H_0$ be defined by
\begin{equation}
\tilde H_0u=-\frac {\al_M}{24}e^{-\tfrac{(m+1)}2t}\Big(e^{\tfrac{(m-3)}2t}u'\Big)'+2\al_M^{-1} e^{4t}u.
\end{equation}
Then, we first want to solve the eigenvalue problems
\begin{equation}
\tilde H_0u=\lam u
\end{equation}
in an appropriate function space.
\bd
For $p=\tfrac{m-3}2$ define $\mc H_0$ as the completion of $C^\un_c(\R[],\Cc)$ with respect to the norm
\begin{equation}
\norm u^2=\int_{\R[]}\abs u^2e^{(p+2)t}
\end{equation}
and  $\tilde{\mc H}_0$ as the completion of $C^\un_c(\R[],\Cc)$ with respect to the norm
\begin{equation}
\norm u^2_1=\int_{\R[]}\{\abs{\dot u}^2e^{pt} +\abs u^2e^{(p+6)t}\}.
\end{equation}
\ed

\bl
The norm $\norm\cdot$ is compact relative to $\norm\cdot_1$.
\el
\bp
Let $u_k\in\tilde{\mc H}_0$ be a sequence converging weakly to zero, then we have to prove
\begin{equation}
\lim\norm{u_k}=0.
\end{equation}

Let $I=(a,b)$ be any bounded interval and $\chi=\chi_I$ be its characteristic function, then
\begin{equation}
\lim\norm{u_k\chi_I}=0,
\end{equation}
in view of the Sobolev embedding theorem saying that the embedding
\begin{equation}
H^{1,2}(I)\hra L^2(I)
\end{equation}
is compact.

Thus, we only have to prove
\begin{equation}\lae{11.25}
\limsup\int_b^\un\abs{u_k}^2e^{(p+2)t}\le \e(b),
\end{equation}
where
\begin{equation}\lae{11.26}
\lim_{b\ra\un}\e(b)=0,
\end{equation}
and a similar estimate in $(-\un,b)$, $b<<-1$,
\begin{equation}\lae{11.27}
\limsup\int^b_{-\un}\abs{u_k}^2e^{(p+2)t}\le \e(b).
\end{equation}

Let us first prove \re{11.25}, which is almost trivial. From 
\begin{equation}
\norm{u_k}_1\le c\qq\A\,k
\end{equation}
we deduce
\begin{equation}
\int_b^\un\abs{u_k}^2e^{(p+2)t}\le e^{-4b}\int_b^\un\abs{u_k}e^{(p+6)t}\le ce^{-4b}\equiv\e(b),
\end{equation}
which implies \re{11.26}. 

The proof of \re{11.27} is a bit more delicate. First, we make a change of variables setting
\begin{equation}
\tau=-t
\end{equation}
such that the crucial estimate for $u_k=u_k(\tau)$ is
\begin{equation}\lae{11.31}
\limsup\int_b^\un\abs{u_k}^2e^{-(p+2)\tau}\le \e(b).
\end{equation}

Replacing $u_k$ by
\begin{equation}
u_k\h,
\end{equation}
where $\h$ is a cut-off function, we may assume without loss of generality that
\begin{equation}\lae{11.33}
\supp u_k\su (\tau_1,\un), \qq\tau_1>3.
\end{equation}
We then only use the estimate
\begin{equation}\lae{11.34}
\int_0^\un \abs{\dot u_k}^2e^{-p\tau}\le c\qq\A\,k
\end{equation}
and the Hardy-Littlewood inequality
\begin{equation}\lae{11.35}
\int_0^\un\abs u^2\tau^{-\s}\le \Big(\frac2{\abs{\s-1}}\Big)^2\int_0^\un\abs{\dot u}^2\tau^{(-\s+2)},
\end{equation}
which is valid for all $u\in C^\un_c(\R[]_+)$ and all $1\ne \s\in\R[]$, \cf \cite[Theorem 3.30]{hardy:book}.

We distinguish two cases.
\bh
Case \tup{1:} $p=0$.
\eh

Then, we may choose in \re{11.35} $\s=2$ and $u=u_k$ to deduce
\begin{equation}
\int_0^\un\abs{u_k}^2\tau^{-2}\le 4\int_0^\un\abs{\dot u_k}^2\le 4 c,
\end{equation}
and we conclude further
\begin{equation}
\int_b^\un\abs{u_k}^2e^{-2\tau}\le b^2e^{-2b}\int_b^\un\abs{u_k}^2\tau^{-2},
\end{equation}
if $b>1$, hence the result.

\bh
Case \tup{2:} $p\ne 0$
\eh

If $p\ne0$, we employ another variable transformation
\begin{equation}
r=e^\tau,
\end{equation}
such that
\begin{equation}
\frac d{d\tau}u\equiv\dot u=\frac d{dr}u e^\tau\equiv u'e ^\tau,
\end{equation}
and we infer
\begin{equation}
\int_0^\un\abs{u'_k}^2r^{(1-p)}=\int_0^\un\abs{\dot u_k}^2e^{-p\tau}\le c,
\end{equation}
in view of  \re{11.33} and \re{11.34}.

Thus, we may apply the Hardy-Littlewood inequality with
\begin{equation}
\s=p+1
\end{equation}
to derive
\begin{equation}
\int_{r_0}^\un \abs{u_k}^2r^{-(p+3)}\le r_0^{-2}\int_{r_0}^\un\abs{u_k}^2r^{-(p+1)}\le c r_0^{-2}=\e(r_0),\q r_0>1,
\end{equation}
where we used \re{11.33}, completing the proof of the lemma.
\ep

Let $\spd\cdot\cdot$ be the scalar product
\begin{equation}
\spd uv=\int_{\R[]}u\bar ve^{pt}
\end{equation}
in $\mc H_0$ and
\begin{equation}
a(u,v)=\spd{\tilde H_0u}v=\int_{\R[]}\Big\{\frac{\al_M}{24}\dot u\dot{\bar v}+ 2\al_M^{-1}u\bar ve^{(p+6)t}\Big\}\q\A\, u,v\in \tilde{\mc H}_0,
\end{equation}
then, by applying the general variational eigenvalue principle, we obtain:
\bt
There exists a sequence of normalized eigenfunctions $\tilde u_i$ with strictly positive eigenvalues $\tilde\lam_i$ with finite multiplicities such that
\begin{equation}
0<\tilde\lam_i\le \tilde\lam_{i+1}\q\wed\q \lim\tilde\lam_i=\un,
\end{equation}
\begin{equation}\lae{11.47}
a(\tilde u_i,v)=\tilde\lam_i\spd{\tilde u_i}v\qq\A\, v\in \tilde{\mc H}_0,
\end{equation}
and
\begin{equation}
a(\tilde u_i,\tilde u_j)=\spd{u_i}{u_j}=0\qq\A\, i\ne j.
\end{equation}
Define the linear operator $\tilde H$ by
\begin{equation}
D(\tilde H)=\langle(\tilde u_i)_{i\in\N}\rangle\q\wed\q \tilde H\tilde u_i=\tilde \lam_i\tilde u_i\q\A\, i,
\end{equation}
then $\tilde H$ is densely defined in $\mc H_0$, symmetric, essentially self-adjoint and
\begin{equation}
a(u,v)=\spd{\tilde H u}v\qq\A\, u,v\in D(\tilde H).
\end{equation}

Moreover,  there holds
\begin{equation}\lae{11.50}
\tilde u_i\in C^\un (\R[],\Cc)
\end{equation}
and
\begin{equation}\lae{11.51}
\tilde H_0 \tilde u_i=\tilde H u_i=\tilde\lam_i\tilde u_i.
\end{equation}
\et
\bp
We only have to prove \re{11.50} and \re{11.51}, since the proof of the other statements is identical to the proof of  \frt{10.2}. 

From \re{11.47} we immediately deduce
\begin{equation}
\tilde H_0\tilde u_i=\tilde\lam_i\tilde u_i
\end{equation}
in the distributional sense, hence \re{11.50} is valid, which in turn implies \re{11.51}.
\ep

An immediate consequence of the preceding result is:
\bt
Let $\mu>0$, then the pairs $(\tilde u_i,\lam_i)$ represent a complete set of eigenfunctions with eigenvalues
\begin{equation}
\lam_i=\tilde\lam_i\mu^{-1}
\end{equation}
for the eigenvalue problems 
\begin{equation}
\tilde H_0u=\lam\mu u.
\end{equation}
The rescaled functions
\begin{equation}
u_i(t)=\tilde u_i(t-\tfrac12\log\lam_i)
\end{equation}
then satisfy
\begin{equation}
-\frac {\al_M}{24}e^{-\tfrac{(m+1)}2t}\Big(e^{\tfrac{(m-3)}2t}u_i'\Big)'+2\al_M^{-1}\lam_i^{-3} e^{4t}u_i=\mu u_i,
\end{equation}
or, if we set
\begin{equation}
\Lam_i=-\lam_i^{-3},
\end{equation}
\begin{equation}
-\frac {\al_M}{24}e^{-\tfrac{(m+1)}2t}\Big(e^{\tfrac{(m-3)}2t}u_i'\Big)'-2\al_M^{-1}\Lam_i e^{4t}u_i=\mu u_i.
\end{equation}
\et

We can now prove the spectral resolution of the Wheeler-DeWitt equation. Let $(\mu,\f)$ \resp $(\lam,\tilde u)$ be a solution of
\begin{equation}
\hat H_1\f=\mu\f
\end{equation}
\resp
\begin{equation}
\tilde H_0\tilde u=\lam\mu \tilde u,
\end{equation}
and set
\begin{equation}
\tilde\Psi=\tilde u\otimes\f,
\end{equation}
then
\begin{equation}
\tilde H_0\tilde\Psi =\lam \hat H_1\tilde\Psi,
\end{equation}
or equivalently, in view of the preceding theorem,
\begin{equation}
H_0\Psi-\hat H_1\Psi=0,
\end{equation}
where
\begin{equation}
\Psi=u\otimes \f,
\end{equation}
\begin{equation}
u(t)=\tilde u(t-\tfrac12\log \lam),
\end{equation}
\begin{equation}
H_0\Psi=-\frac {\al_M}{24}e^{-\tfrac{(m+1)}2t}\Big(e^{\tfrac{(m-3)}2t}\Psi'\Big)'-2\al_M^{-1}\Lam e^{4t}\Psi,
\end{equation}
and
\begin{equation}
\Lam=-\lam^{-3}.
\end{equation}

One easily checks that $\Psi$ belongs to
\begin{equation}
\tilde{\mc H}_0\otimes\tilde{\mc H}_1\su \mc H_0\otimes \mc H_1,
\end{equation}
\cf the corresponding considerations in \cite[section 3]{cg:qfriedman-ym2}.

Let $\tilde u_i$ \resp $\f_j$ be the eigenfunctions of $\tilde H_0$ \resp $\hat H_1$, then
\begin{equation}
\tilde\Psi_{ij}=\tilde u_j\otimes\f_j
\end{equation}
form a complete set of eigenfunctions in $\mc H_0\otimes\mc H_1$ of the linear operator
\begin{equation}
H=\tilde H_0\hat H_1^{-1}=\hat H_1^{-1}\tilde H_0,
\end{equation}
such that
\begin{equation}
H\tilde\Psi_{ij}=\lam_{ij}\tilde\Psi_{ij}=\lam_i\mu_j^{-1}\tilde\Psi_{ij},
\end{equation}
where
\begin{equation}
D(H)=\langle(\tilde\Psi_{ij})_{(i,j)\in\N\times\N}\rangle.
\end{equation}

The rescaled functions
\begin{equation}
\Psi(t,\cdot)=\tilde\Psi(t-\tfrac12\log\lam_{ij},\cdot)
\end{equation}
are solutions of the Wheeler-DeWitt equation with cosmological constant
\begin{equation}
\Lam_{ij}=-\lam_{ij}^{-3}.
\end{equation}
\br\lar{11.5}
$H$ is essentially self-adjoint in $\mc H_0\otimes\mc H_1$ and we consider it to be the Hamiltonian associated with the physical system defined by the interaction of gravity with the matter fields. The properly rescaled eigenfunctions $\Psi_{ij}$ are solutions of the Wheeler-DeWitt equation. We refer to \cite[section 3]{cg:qfriedman-ym2}, where these connections have been explained and proved in greater detail.

The wave functions $\Psi$ are maps from
\begin{equation}
\Psi:\R[4n+10]\ra \mc F_1\otimes\mc F_2
\end{equation}
and in general the eigenstates $\Psi$ cannot be written as simple products
\begin{equation}
\Psi=u\h,
\end{equation}
such that
\begin{equation}
\h\in \mc F_1\otimes\mc F_2\q\wed\q u(x)\in \Cc\qq\A\, x\in \R[4n+10].
\end{equation}
Thus, in general it makes no sense specifying a fermion $\h$ and looking for an eigenfunction $\Psi$ satisfying
\begin{equation}
R(\Psi)\su \langle\h\rangle.
\end{equation}
\er

%\backmatter

\bibliographystyle{hamsplain}
%\bibliography{mrabbrev,publications}
\providecommand{\bysame}{\leavevmode\hbox to3em{\hrulefill}\thinspace}
\providecommand{\href}[2]{#2}

%\listoffigures

%\cleardoublepage

%\thispagestyle{empty}
%\closegraphsfile
\end{document}